\documentclass[english]{article}

\usepackage[T1]{fontenc}
\usepackage[latin9]{inputenc}
\usepackage{geometry}
\usepackage{color}
\usepackage{float}
\usepackage{amsmath}
\usepackage{graphicx}
\usepackage[numbers]{natbib}

\makeatletter
\newcommand{\lyxaddress}[1]{
	\par {\raggedright #1
	\vspace{1.4em}
	\noindent\par}
}

\@ifundefined{date}{}{\date{}}
\makeatother

\usepackage{babel}
\begin{document}
\title{Traction chain networks: Insights beyond force chain networks for
non-spherical particle systems}
\author{{\small{}Daniel N. Wilke}$^{1}$}
\maketitle

\lyxaddress{\noindent $^{1}$\textcolor{black}{Department of Mechanical and Aeronautical
Engineering, University of Pretoria, South Africa. }}

\lyxaddress{\noindent \textcolor{black}{E-mail: nico.wilke@up.ac.za}}
\begin{abstract}
Force chain networks are generally applied in granular materials to
gain insight into inter-particle granular contact. For conservative
spherical particle systems, i.e. frictionless and undamped, force
chains are information complete due to symmetries resulting from isotropy
and constant curvature of a sphere. In fact, for conservative spherical
particle systems, given the geometry and material, the force chain
network uniquely defines the contact state that includes elastic forces,
penetration distance, overlap volume, contact areas and contact pressures
in a particle system. This is, however, not the case for conservative
non-spherical particle systems. The reason is that a force chain network
is not sufficient to uniquely define the contact state in a conservative
non-spherical particle system. Additional information is required
to define the contact state of non-spherical granular systems. Traction
chain networks are proposed to complement force chain networks for
the improved quantification of the state of contact of a granular
system. 
\end{abstract}

\section{\textcolor{black}{Introduction}}

Particles under dynamical loading continuously experience a change
in forces and accelerations to which the particles are subjected,
while in static particle systems, these loadings remain fixed. Particle
systems are loaded as the result of body forces and surface tractions.
The result is a complex response of particles spatially and temporally
that manifest as flow, transitional and solid response over space
and through time. 

The distinction between body forces and surface tractions differentiates
between loadings that act at each point in a body, i.e. body forces,
from loadings that only act only on the surface of a particle, i.e.
surface tractions. Gravity and inertial forces represent body forces,
while pseudo body forces described by non-inertial reference frames
include centrifugal, Euler and Coriolis forces. In turn, surface tractions
are due to interactions between particles and the environment that
may include geometrical objects or fluids and between particles. Surface
tractions resulting from direct contact between particles include
normal contact, sliding contact due to friction, interlocking for
non-convex systems \citep{wilke2016computing,Wilke2017,GOVENDER2018_NONCONVEX},
short-range cohesive and adhesive interactions.

Spherical particle contact has been studied for a variety of cases.
These include the classical frictionless and small strain solution
of Hertz \citep{Hertz1882}, friction and adhesive (e.g. van der Waals)
forces with the work of Johnson, Kendall and Roberts (JKR) that generalised
Hertz\textquoteright{} solution for elastic spheres \citep{jkr1971}.
Although contact between spheres is well understood, spherical particle
systems exhibit complex behaviour \citep{Vallejo2005}. 

Force chain networks form an integral component to study and quantify
the resulting complex behaviour \citep{Mueth_1998}. They are heterogeneous
structures that are ubiquitous in granular material research as they
quantify aspects of load, acoustic and thermal paths through granular
materials. Force chain networks quantify contact forces between interacting
particles, distinct from stress networks that quantify the distribution
of the average particle stresses in granular media \citep{Luding1997}.
Ironically, despite the prevalence of force chain networks, there
is no agreement on a quantitative description of force chains, numerous
approaches having been proposed as outlined by \citep{Bassett_2015}.
Despite this force chain networks aim to quantify the network of contact
forces between particles.

Experimental attempts to visualise force chain networks include photoelasticity,
acoustically \citep{Bassett_2015}, thermal conduction \citep{Grinchuk2018}
and electrical conduction \citep{Falcon2004}. In particular, isochromatic
photoelasticity visualises differences in principal stress directions.
Acoustic transmission paths highlight paths with matched contact impendence.
Thermal conduction visualises thermal transmission paths that mainly
depend on the effective contact area. Electrical conduction paths
are observed after an initial insulating state changes to conducting
state when the DC Branly threshold is exceeded. Physical transmission
paths are highly correlated with interparticle contact forces for
spherical particle systems. This stems from the fact that the state
of inter-particle contact is completely and uniquely defined by a
force chain network for a conservative spherical particle system.

\begin{figure}[H]
\begin{centering}
\includegraphics[scale=0.25]{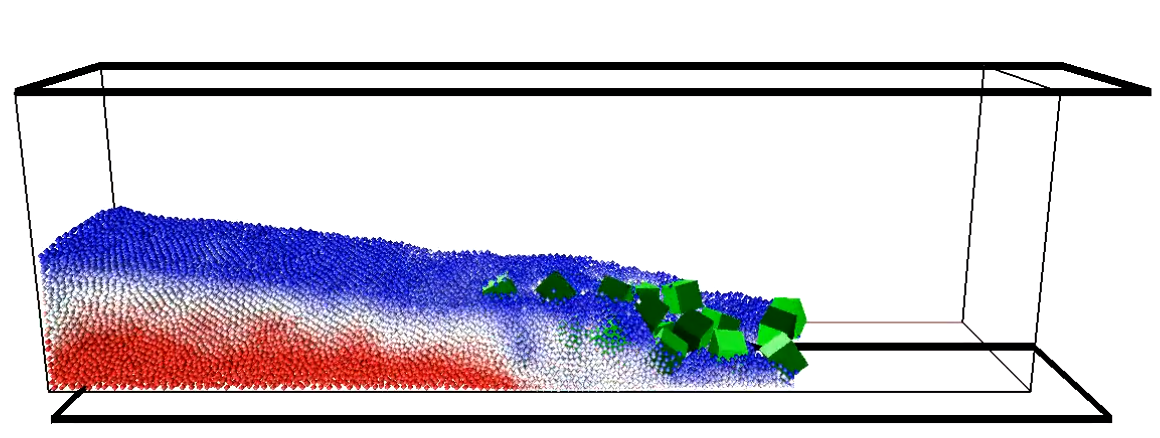}
\par\end{centering}
\caption{Resulting complex surface tractions due to interaction between particles
(light green) and its environment that includes fluids. The fluid
colors indicate normalised pressure (red high and blue low) \citep{joubert2020}.
\label{fig:demsph}}
\end{figure}

However, even for conservative non-spherical particle systems, this
is not the case. Physical transmission paths require additional information
to uniquely define the contact state of force and pressure. As a result,
force chain networks give only a partial view of the state of a network.
Further, non-spherical particle contact is much less understood. Contact
between angular particles may result in localised tractions, e.g.
near-point contact or near edge contacts or distribute over an entire
face that may result in the same contact forces but yield distinct
surface tractions. Additional complex surface tractions may result
from interactions with the environment, fluids and gases, as shown
in Figure \ref{fig:demsph}, for fluid modelled using smoothed-particle
hydrodynamics (SPH) interacting with cuboid particles \citep{joubert2020}.

This paper extends the notion of informative networks beyond inter-particle
contact forces to also consider inter-particle contact tractions.
This subtle but critical distinction is particularly important when
considering contact between non-spherical particles. Inter-particle
contact tractions complement force chain networks to provide a complete
picture of particle interactions and mechanisms that may contribute
to define the contact state that may help define damage and failure.
Thus, in addition to contact forces, contact tractions offer key insights
required to evolve our understanding and study of non-spherical particle
interactions, degradation and ultimately breakage. The first step
towards this larger study is to quantify and visualise traction chain
networks and force chain networks, which is the  focus of this study.

This paper outlines the required background knowledge on contact mechanics,
macroscopic resultants and tractions in Section \ref{sec:Contact-mechanics},
followed by a compact outline of force and traction chain networks
in Section \ref{sec:Force-chain-network}. Next, force and traction
chains for spherical particle systems are explored in Section \ref{subsec:Spherical-particle-particle}.
This is followed by force and traction chains for polyhedral particle
systems in Section \ref{subsec:Polyhedral-particle-particle-con}.
Non-spherical particle systems considered in this study is limited
to polyhedral particle systems without the loss of generality. Finally,
conclusions and future work are offered in Section \ref{sec:Conclusion}.

\section{Background\label{sec:Contact-mechanics}}

\subsection{Spherical contact mechanics}

\subsubsection{Hertz-Mindlin model}

Consider the classical solution for non-adhesive frictionless elastic
contact between two spheres, $i$ and $j$, with radii $R_{i}$ and
$R_{j}$ is equated with contacting between a sphere with effective
radius $R$ computed from 

\begin{equation}
\frac{1}{R}=\frac{1}{R_{i}}+\frac{1}{R_{j}},
\end{equation}

and a half-space or plane. The radius of the contact area, $a$, between
an elastic sphere with effective radius $R$ that penetrates a half-space
or plane by a distance, $d$, is given by

\begin{equation}
a=\sqrt{Rd},
\end{equation}

with contact Area, $A(d)$, being linearly related to the penetration
distance, $d$, following

\begin{equation}
A=\pi Rd.
\end{equation}

In turn, the applied force $F$ is related to the penetration distance,
$d$, by

\begin{equation}
F=\frac{4}{3}E^{*}R^{\frac{1}{2}}d^{\frac{3}{2}},
\end{equation}

with effective Young's modulus

\begin{equation}
\frac{1}{E^{*}}=\frac{1-\nu_{s}^{2}}{E_{s}}+\frac{1-\nu_{p}^{2}}{E_{p}},
\end{equation}

computed from the individual Young's modulus of the sphere (subscript
$s$) and plane (subscript $p$). The radial variation in contact
pressure is expressed by

\begin{equation}
p(r)=p_{max}(1-\frac{r^{2}}{a^{2}})^{\frac{1}{2}},
\end{equation}

where the maximum pressure $p_{max}$ is given by

\begin{equation}
p_{max}=\frac{3}{2}\frac{F}{\pi a^{2}}.
\end{equation}

The radius of the contact area is related to the applied load $F$
by

\begin{equation}
a^{3}=\frac{3}{4}\frac{FR}{E^{*}},
\end{equation}

Consequently, the penetration distance is related to the maximum contact
pressure by

\begin{equation}
d=\frac{3}{2}\frac{F}{\pi p_{max}R}.
\end{equation}

The contact volume overlap is given by 

\begin{equation}
V=\frac{\pi d^{2}}{3}(3R-d).
\end{equation}

\begin{figure}[H]
\begin{centering}
\includegraphics[scale=0.25]{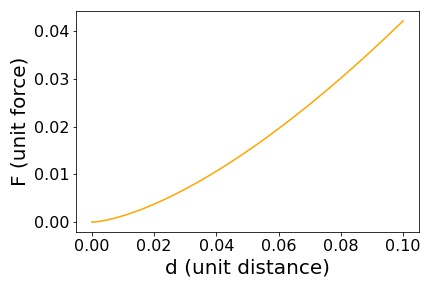}\includegraphics[scale=0.25]{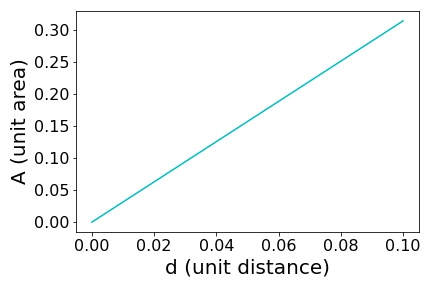}\includegraphics[scale=0.25]{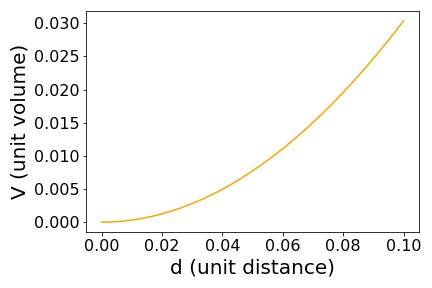}\includegraphics[scale=0.25]{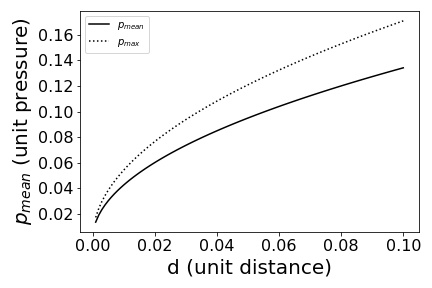}
\par\end{centering}
\begin{centering}
\includegraphics[scale=0.4]{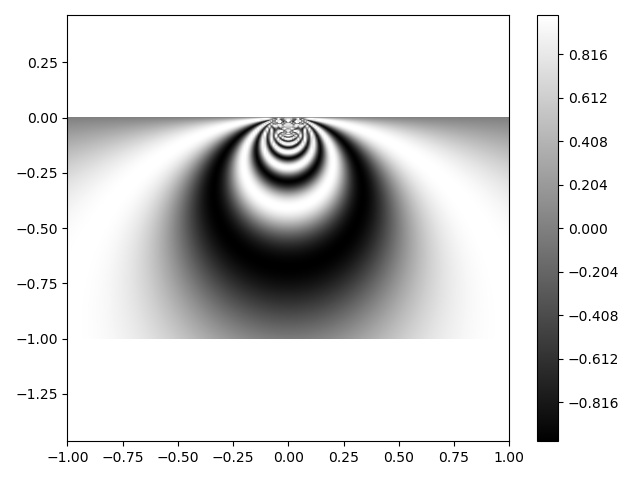}
\par\end{centering}
\caption{Top left to right: Contact force magnitude ($F$), contact area ($A$),
contact volume ($V$) and average pressure $P_{mean}$ as a function
of penetration distance for a frictionless sphere. Centre bottom:
maximum shear stress cosine inside an infinite plane, representing
a bifringe pattern.\label{fig:Left:-unit-area}}
\end{figure}

For a Poission's ratio of $\nu=0.33,$ the maximum shear stress occurs
at $z\approx0.49a.$ Figure \ref{fig:Left:-unit-area} show the contact
force, contact area, contact volume and mean pressure relationships
w.r.t. a penetration distance of up to a 10\% between a frictionless
sphere and an infinite plate (left), and the maxmimum shear stress
cosine distribution inside the infinite plate. Consequently, $F(d)$,
$V(d)$, $A(d)$ and $P_{mean}(d)$, indicating that $d$ uniquely
defines the state of $F$, $V$, $A$ and $P_{mean}$.

For circular grains subject to NC contacts, the stress tensor \citep{Luding1997}
is estimated from

\begin{equation}
\sigma_{ij}=\frac{1}{V}\sum_{c=1}^{NC}\mathbf{l_{i}\cdot F_{j},}
\end{equation}

with $\textbf{l}$ the branch vector of particles in contact and $\textbf{F}$
the contact forces between particles.

Hence, for conservative contact, i.e. frictionless and undamped contact
we find that $F$, $A$, V, $p_{mean}$ and $p_{max}$ are only a
function penetration distance given the material properties and radii.
As a result force chains are informative and fundamental to quantifying
spherical contact mechanics. 

\subsubsection{Linear-spring dashpot model}

The normal and tangential forces are often modelled by a linear-spring
dashpot (LSD) model \citep{POSCHEL2005}. Consider two particles in
contact, particle $i$ and $j$ (indicated as subscripts with variables).
Their center of mass (COM) positions are given by $\boldsymbol{x}_{i}$
and $\boldsymbol{x}_{j}$ and the contact points relative to the COMs
represented by $\boldsymbol{r}_{i}$ and $\boldsymbol{r}_{j}$. The
linear and angular velocities are given by $\boldsymbol{v}_{i}$and
$\boldsymbol{v}_{j}$; and $\boldsymbol{\omega}_{i}$ and $\boldsymbol{\omega}_{j}$,
for the two particles, respectively. The relative velocity $\boldsymbol{v}_{ij}$
at the contact points is given $\boldsymbol{v}_{ij}=\boldsymbol{v}_{j}-\boldsymbol{v_{i}}+\boldsymbol{\boldsymbol{\omega}}_{j}\times\boldsymbol{r_{j}}-\boldsymbol{\omega}_{i}\times\boldsymbol{r_{i}}$.
The unit vectors $\boldsymbol{u}_{ij}=\frac{(\mathbf{x_{j}-x_{i}})}{||\boldsymbol{x}_{j}-\boldsymbol{x}_{i}||_{2}}$
and $\boldsymbol{t}_{ij}=\frac{\boldsymbol{v}_{ij}}{||\boldsymbol{v}_{ij}||_{2}}-\boldsymbol{u}_{ij}$
define the normal and tangential directions. This allows for the decomposition
of the velocity into normal and tangential components given by $\boldsymbol{v}_{ij}^{n}=(\boldsymbol{v}_{ij}\cdot\boldsymbol{u}_{ij})\boldsymbol{u}_{ij}$
and $\boldsymbol{v}_{ij}^{t}=(\boldsymbol{v}_{ij}\cdot\boldsymbol{t}_{ij})\boldsymbol{t}_{ij}$.
The normal and tangential forces between the two particles are computed
by summing the spring (s) and dashpot (d) contributions in the normal
and tangential directions

\begin{equation}
\boldsymbol{F}_{ij}^{u}=\boldsymbol{F_{ij}^{u,s}+F_{ij}^{u,d},}
\end{equation}

\begin{equation}
\boldsymbol{F}_{ij}^{t}=\boldsymbol{F_{ij}^{t,s}+F_{ij}^{t,d}}.
\end{equation}

Given the overlap distance, $d,$ in the normal and, $\delta$ in
the tangential directions, we obtain 

\begin{equation}
\boldsymbol{F}_{ij}^{u,spr}=-k_{n}d\boldsymbol{u}_{ij},
\end{equation}

\begin{equation}
\boldsymbol{F}_{ij}^{t,spr}=-k_{t}\delta\boldsymbol{t}_{ij},
\end{equation}

for normal and tangential spring stiffness's $k_{n}$ and $k_{t}.$
Consequently, the magnitude of the normal force is given by

\begin{equation}
\|\boldsymbol{F^{u,s}}_{ij}\|=kd.
\end{equation}

Recall, the contact radius is given by $a=\sqrt{Rd}$, which combined
with the former, gives

\begin{equation}
p_{mean}=\frac{kd}{\pi a^{2}}=\frac{kd}{\pi Rd}=\frac{k}{\pi R}.
\end{equation}

As a result, the $p_{mean}$ is constant, which is a direct consequence
of the linear spring model. Hence, when employing a LSD model, we
expect to recover a constant traction magnitude irrespective of the
penetration distance when damping is inactive, and friction ignored.

\subsection{Polyhedral contact mechanics}

\begin{figure}[H]
\begin{centering}
\includegraphics[scale=0.3]{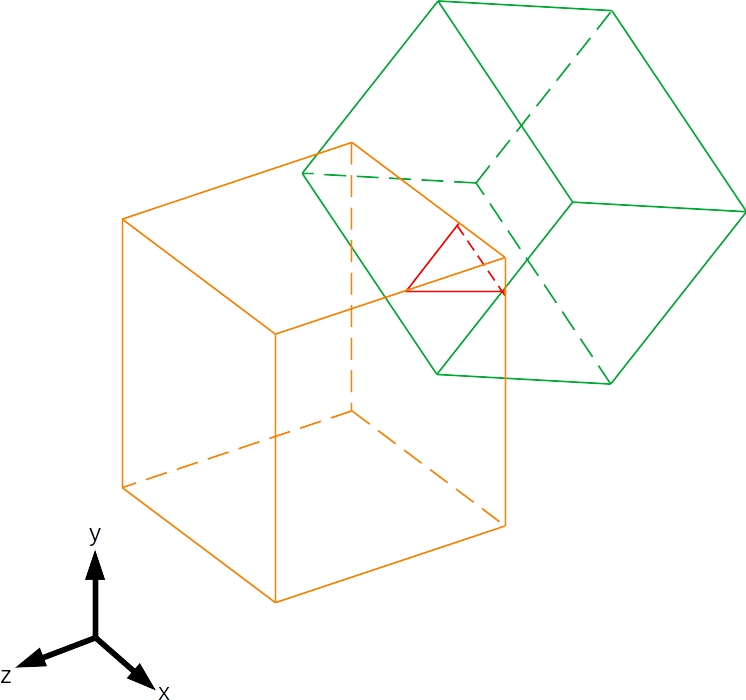} \includegraphics[scale=0.3]{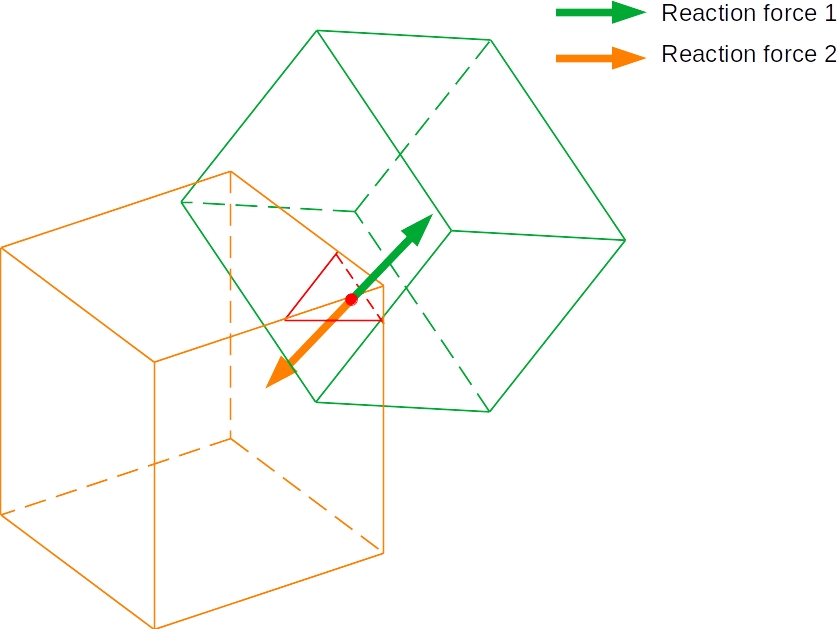}
\includegraphics[scale=0.2]{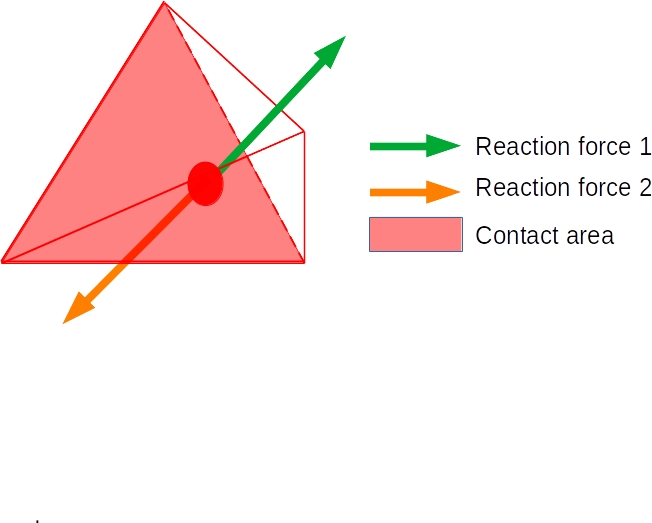}
\par\end{centering}
\caption{Overlap volume-based polyhedral contact between the orange and green
cubes (left) from which the overlap volume (right) is computed. The
green and orange arrows indicate the resultant forces, and the transparent
red surface the contact area. \label{fig:Overlap-volume-based-polyhedral}}
\end{figure}

Non-adhesive frictionless elastic contact between two polyhedra is
visualised in Figure \ref{fig:Overlap-volume-based-polyhedral}. The
overlap volume uniquely resolves the contact force directions. It
avails information to estimate the contact force magnitude, which
can be expressed as a function of $f(d,A,V)$, penetration distanced
$d$, surface area $A$ and volume $V.$ Commercial and research DEM
codes mostly resolve force magnitude as some function of penetration
distance $f(d)$ (as used by Rocky \citep{ROCKY}) or overlap volume
$f(V)$ \citep{GOVENDER2018_NONCONVEX,FENG2021_Volume}. The former
offers computational advantages, while the latter offers continuity
and stability.

\begin{figure}[H]
\begin{centering}
\includegraphics[scale=0.3]{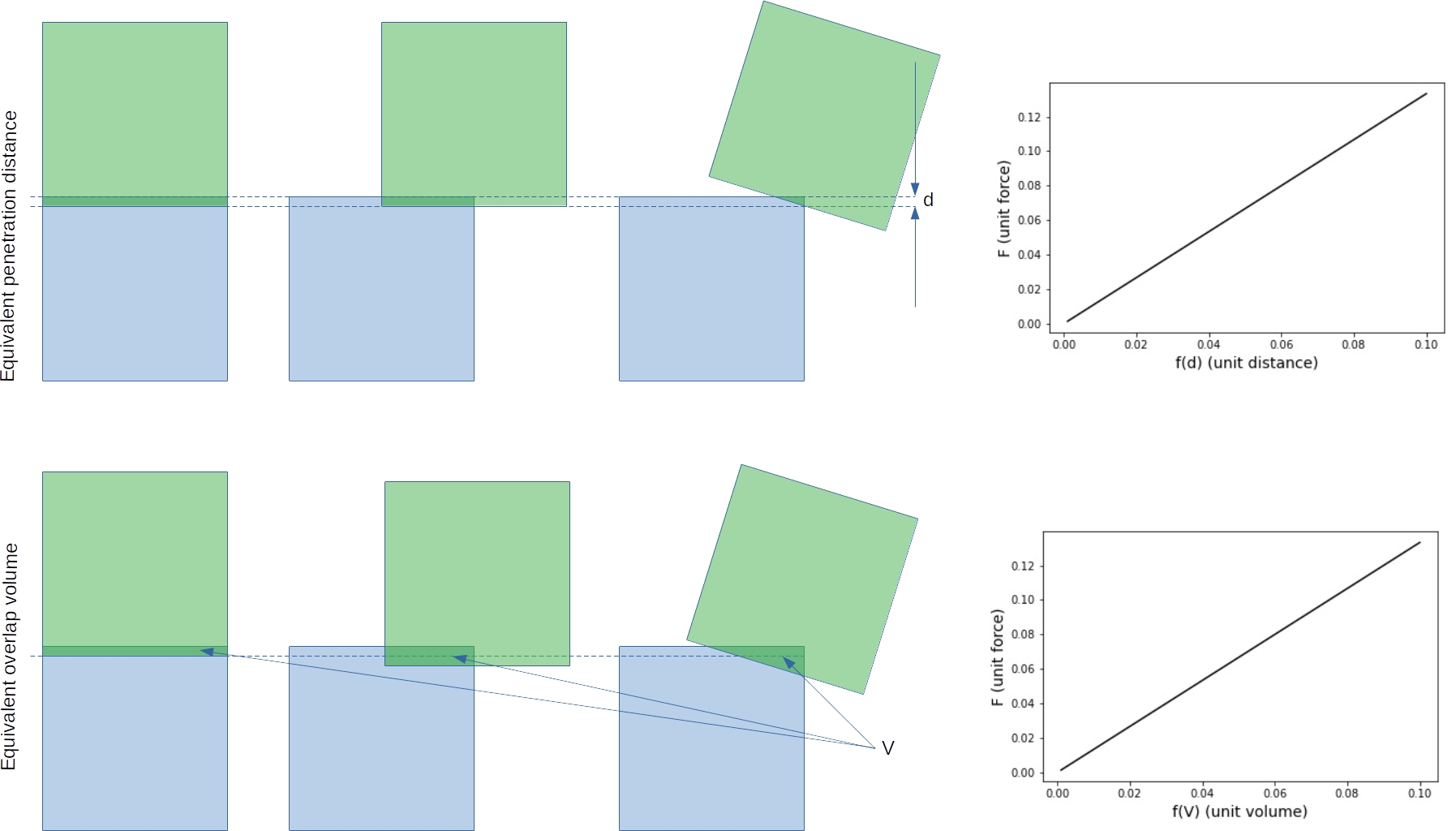}
\par\end{centering}
\caption{Two-dimensional polyhedral contact scenarios for prismatic shapes
of unit thickness (contact overlap is exaggerated for clarity). Top
row represents equal penetration distances, while the bottom row represents
equal overlap volumes. The force distance or volume relationship can
be linear or non-linear, with linear indicated for simplicity. \label{fig:Polyhedral-contact-scenarios}}
\end{figure}

For visual clarity, the main ideas are developed around two-dimensional
polyhedral particle representations. This is done without loss of
generalisation as they also represent three-dimensional prisms in
the out of page direction. For example, consider the two-dimensional
visualisation of several polyhedral contact scenarios and force magnitude
computations depicted in Figure \ref{fig:Polyhedral-contact-scenarios}.
The top row represents contact laws that scale contact force magnitude
only with penetration distance, and the bottom row represents contact
laws that scale only with overlap volume \citep{FENG2021_Volume}.

It is important to note that the computed contact force magnitude
and contact force direction is the same for all three scenarios. However,
the maximum or average pressure is distinct for the three scenarios.
The left-most has the lowest average contact pressure and the highest
average contact pressure for the right-most configurations. As a result,
the contact force $F$, contact area $a$, and contact volume $V$
are not uniquely related by the penetration distance as with contact
between spherical particles.

Therefore, polyhedral and other non-spherical particle contacts require
a fundamental rethinking of the role of force chains as the only approach
to quantify and visualise inter-particle contact. 

\subsection{Force and moment resultants and traction vectors\label{sec:Traction-vector-and}}

\begin{figure}[H]
\begin{centering}
\includegraphics[width=0.4\textwidth]{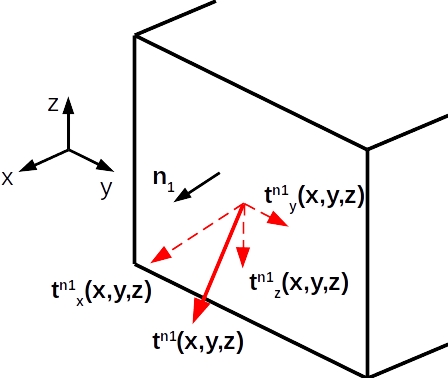}
\par\end{centering}
\caption{Consider the surface with area $A^{\boldsymbol{n_{1}}}$ defined by
outward-pointing unit normal $\boldsymbol{n_{1}}$through which the
traction vector $\boldsymbol{t^{\boldsymbol{n_{1}}}}(y,z)=[t_{x}^{\boldsymbol{n_{1}}},t_{y}^{\boldsymbol{n_{1}}},t_{z}^{\boldsymbol{n_{1}}}]$
acts.}

\label{fig:tractions}
\end{figure}

Computational mechanics defines the resultant surface forces, moments
and torsions to result from tractions acting through surfaces \citep{Holzapfel2002}.
For example, consider the finite surface defined by the unit outward
normal $\boldsymbol{n_{1}}$ with area $A^{\boldsymbol{n_{1}}}$presented
in Figure \ref{fig:tractions}. The unit normal is expressed in the
indicated global coordinate system. Through the surface acts a spatially
varying traction vector $\boldsymbol{t^{\boldsymbol{n_{1}}}}(y,z)=[t_{x}^{\boldsymbol{n_{1}}},t_{y}^{\boldsymbol{n_{1}}},t_{z}^{\boldsymbol{n_{1}}}]$
(force per unit area). The three resultant forces acting on the surface
due to the traction vector are given by
\begin{flushleft}
\begin{align*}
F_{x} & =\int_{A^{\boldsymbol{n}_{1}}}t_{x}^{\boldsymbol{n}_{1}}dA,\\
F_{y} & =\int_{A^{\boldsymbol{n}_{1}}}t_{y}^{\boldsymbol{n}_{1}}dA,\\
F_{z} & =\int_{A^{\boldsymbol{n}_{1}}}t_{z}^{\boldsymbol{n}_{1}}dA,
\end{align*}
the two resultant bending moments are computed from
\par\end{flushleft}

\begin{align*}
M_{y} & =\int_{A^{\boldsymbol{n}_{1}}}t_{x}^{\boldsymbol{n}_{1}}zdA,\\
M_{z} & =\int_{A^{\boldsymbol{n}_{1}}}t_{x}^{\boldsymbol{n}_{1}}ydA,
\end{align*}

and the resultant torsion defined by

\begin{align*}
T_{x} & =\int_{A^{\boldsymbol{n}_{1}}}t_{z}^{\boldsymbol{n}_{1}}ydA-\int_{A^{\boldsymbol{n}_{1}}}t_{y}^{\boldsymbol{n}_{1}}zdA.
\end{align*}

The average traction vector $\boldsymbol{t}_{mean}^{n_{1}}$ due to
load \textbf{$\boldsymbol{F=[F_{x},F_{y},F_{x}]}$ }acting over the
finite surface defined by outward unit normal $\boldsymbol{n_{1}}$
with surface area $A^{\boldsymbol{n_{1}}}$ can be estimated from

\begin{equation}
\boldsymbol{t_{mean}^{n_{1}}=\frac{F}{A^{n_{1}}}}.
\end{equation}

Hence, the traction indicates the intensity of the force acting over
the surface as it has units of force per unit area. For surfaces defined
by normals parallel to the global coordinate system, the tractions
are proportional to the stress components of the stress tensor. As
a result, the stresses acting on the surfaces can readily be derived
from the traction vector.

Note the fundamental role that the area plays in estimating surface
tractions. In contrast, proposed estimates of the particle stress
tensor \citep{Luding1997}, are often independent of the area on which
contact forces act.

\section{Force and traction chain networks\label{sec:Force-chain-network}}

Force chain networks offer insight into contact mechanics that are
informative for spherical particle systems. We show that non-spherical
particle systems require complementary networks to get a more informative
assessment of contact mechanics.

\begin{figure}[H]
\begin{centering}
\includegraphics[scale=0.3]{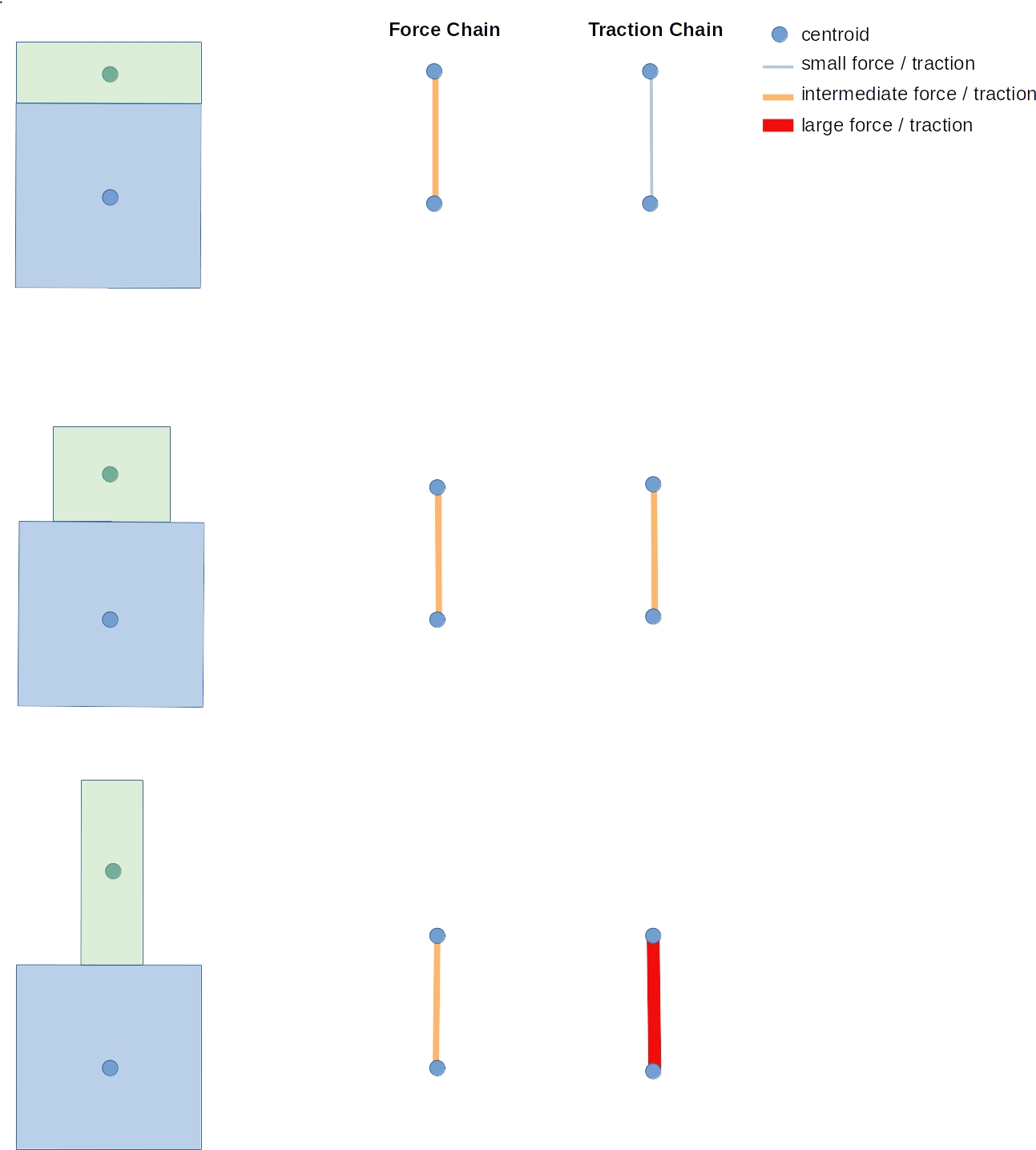}
\par\end{centering}
\caption{Two-dimensional polyhedral contact scenarios with equal mass top polyhedra
under gravity. The result is equivalent contact force magnitudes but
distinct traction magnitudes.\label{fig:equivalent_scenarios}}
\end{figure}

This study proposes the traction chain network to be considered iand
force chain networks to quantify inter-particle contact in particle
systems. Force chain networks and mean particle stress networks offer
complementary viewpoints in particle system interactions. Traction
chain networks complement these networks to highlight surface traction
magnitudes in a particle system. This enables the differentiation
between distinct scenarios that may otherwise seem equivalent, as
shown in Figure \ref{fig:equivalent_scenarios}. For example, consider
three scenarios of binary particle systems of equal mass cuboids.
The force chain network is unable to differentiate between the three
scenarios shown. In turn, the traction chain network offers a complementary
view of contact in the network, clearly showing higher pressure in
the bottom scenario of Figure \ref{fig:equivalent_scenarios}, as
opposed to the middle or top scenarios. Differentiating between these
scenarios is key when considering and quantifying particle damage
and degradation.

\begin{figure}[H]
\begin{centering}
\includegraphics[scale=0.19]{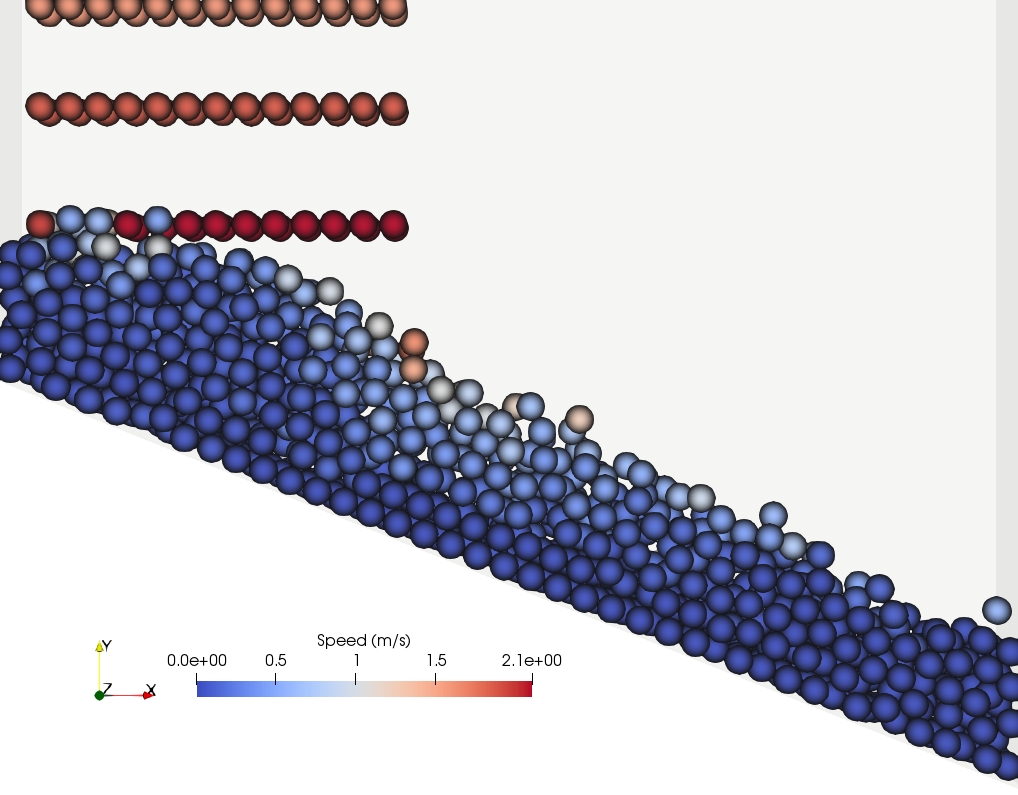} \includegraphics[scale=0.19]{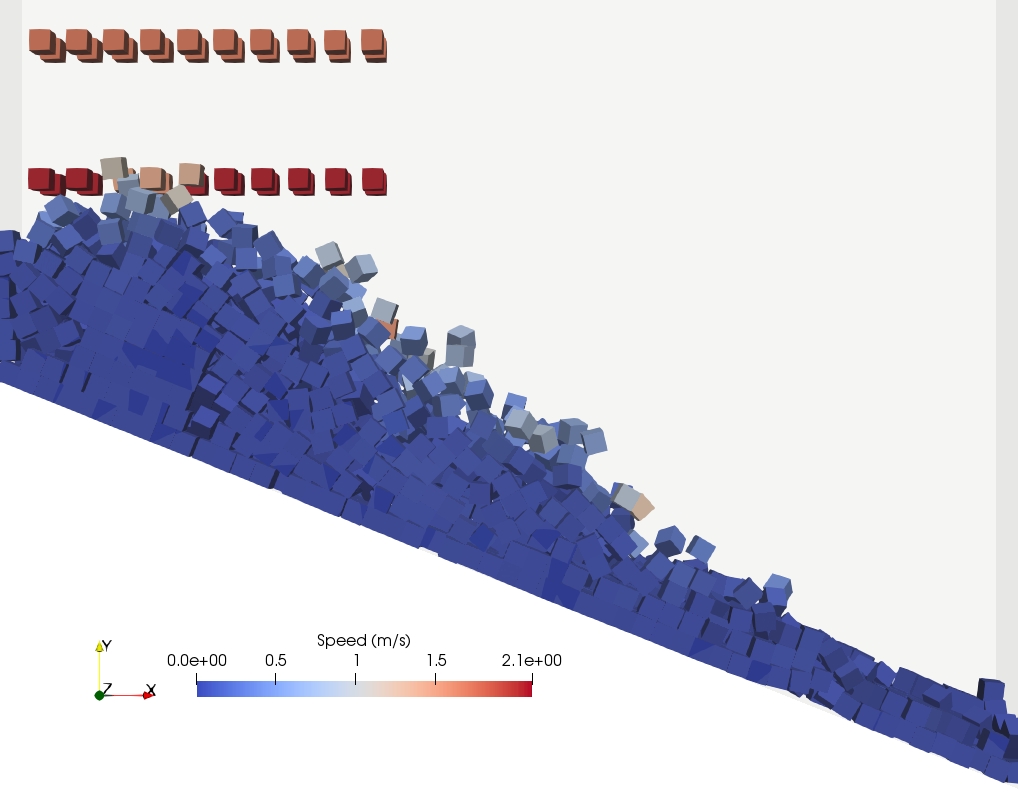}
\par\end{centering}
\begin{centering}
\includegraphics[scale=0.18]{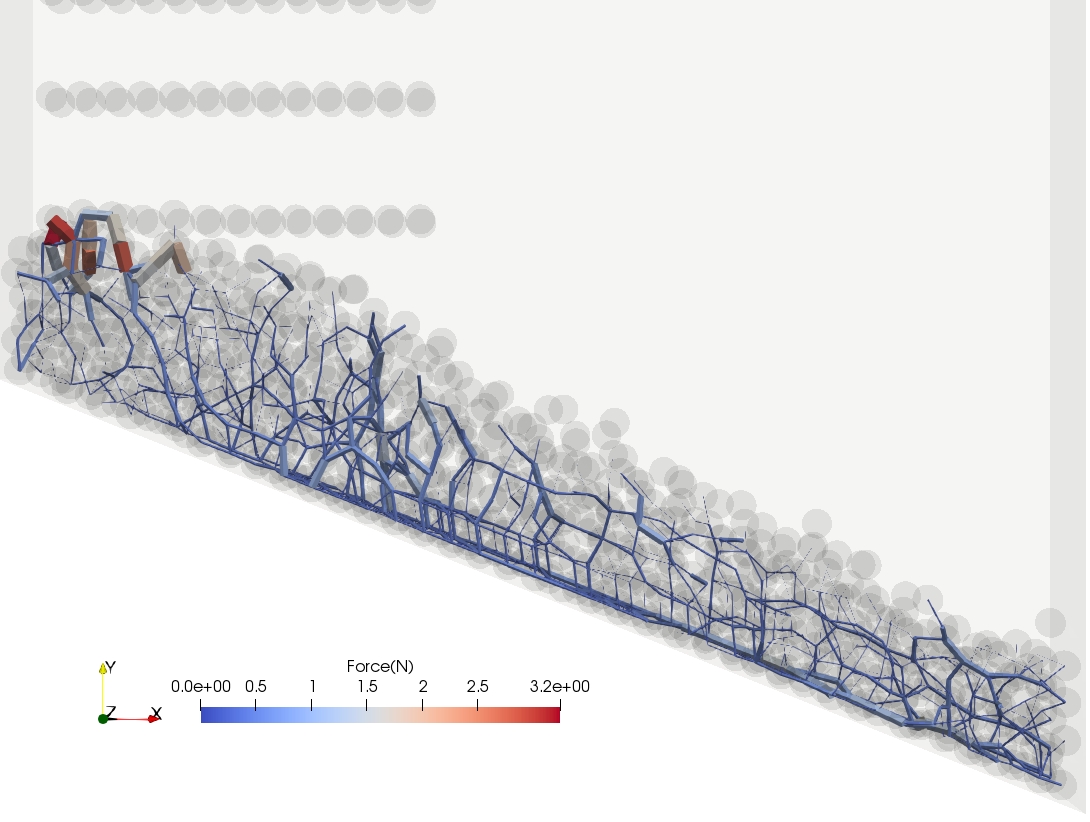}\includegraphics[scale=0.18]{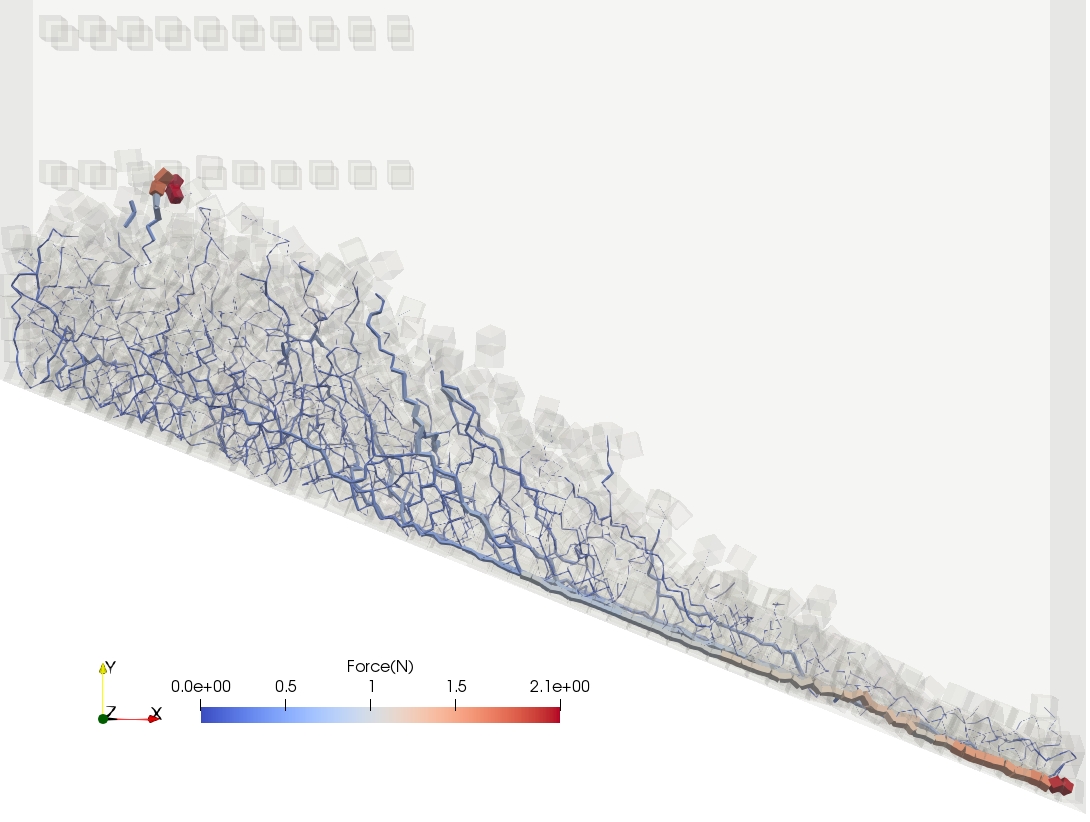}
\par\end{centering}
\begin{centering}
\includegraphics[scale=0.18]{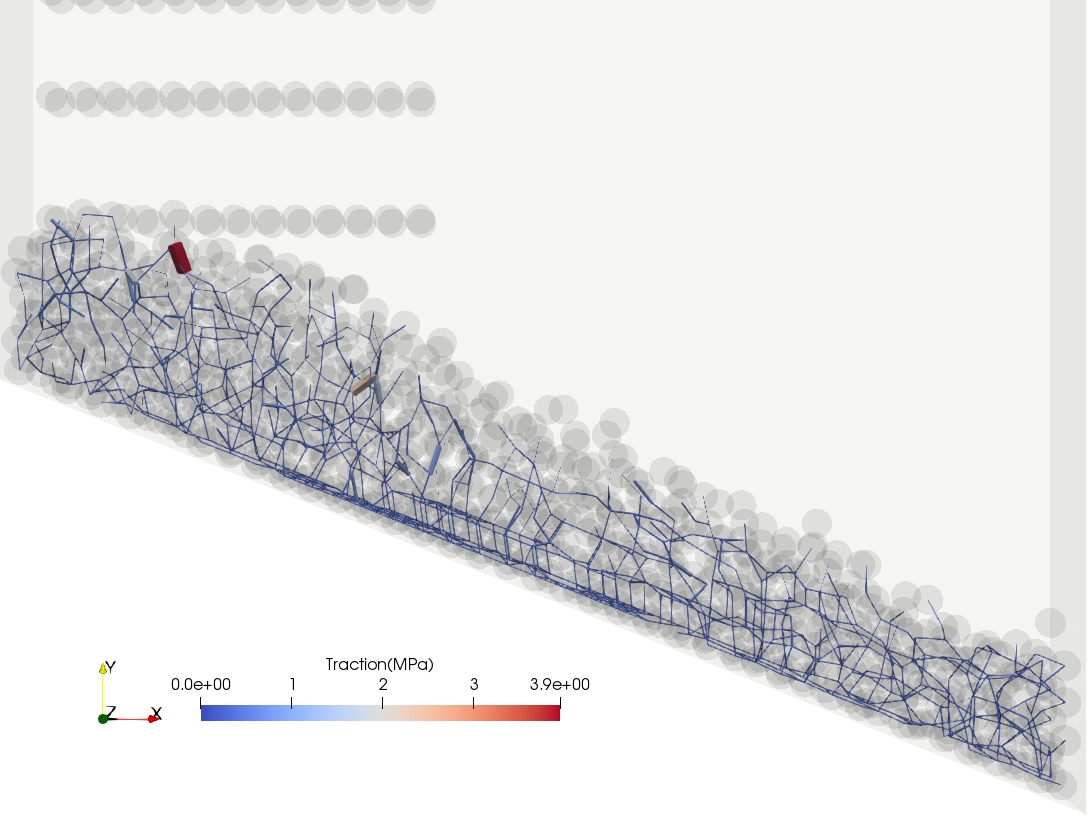}\includegraphics[scale=0.18]{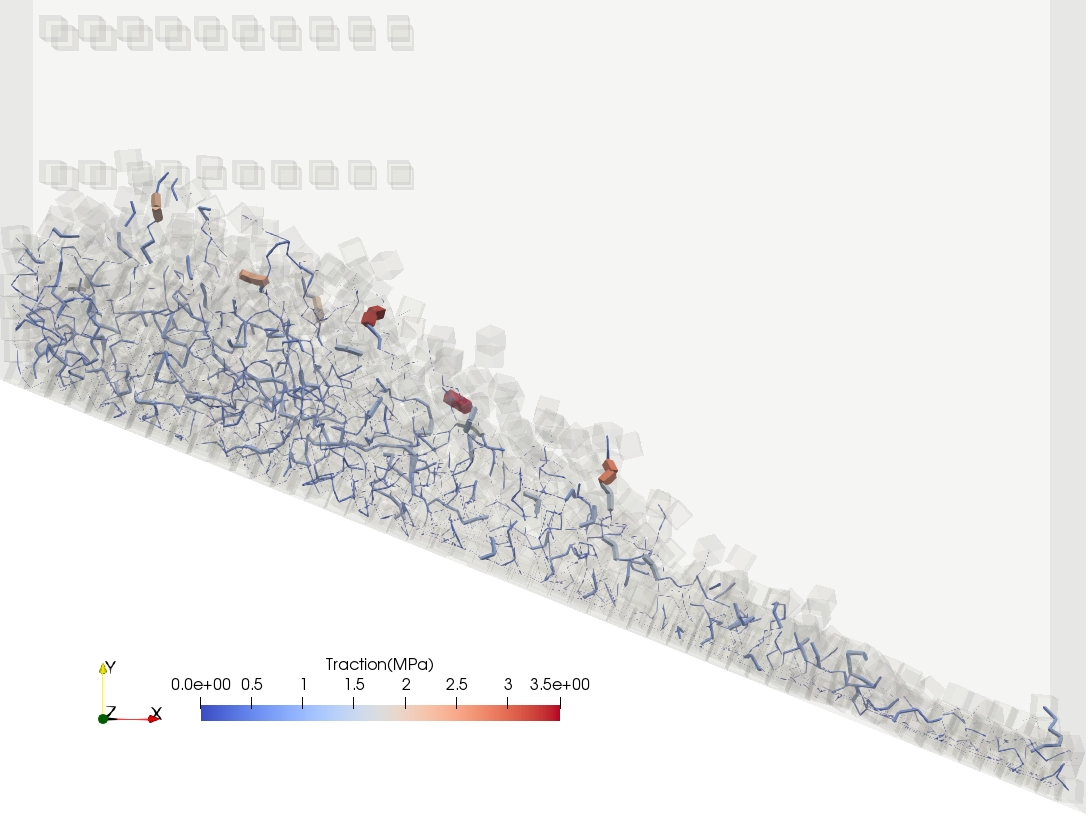}
\par\end{centering}
\caption{Spherical (left) and cubic (right) particles being filled into a slide
box and the resulting force (middle) and traction (bottom) chain networks.}

\label{fig:force_traction_chains}
\end{figure}

Particles experience forces due to particles contacting each other
within a particle system, as shown in Figure \ref{fig:force_traction_chains}
for spherical (left) and cubic (right) particle systems, with each
system containing around 1150 particles.

Body forces load the particles resulting in interactions with the
environment and each other. Particles are often subject to multiple
surface tractions due to inter-particle contact. Several approaches
have been proposed to compute force chains. These include a quasi-linear
approach proposed by \citep{Peters2005} followed by \citep{Arevalo2010}.
Figure \ref{fig:force_traction_chains} visualises the force and traction
chain networks representing the force magnitude between particle centroids
and the contact volume centroid. 

For spherical particle systems, a force chain network is informative.
However, a traction chain network provides additional information
to complement a force chain network when non-conservative systems
(frictional and dissipative particle systems) are considered. 

For conservative particle systems (no friction and undamped), contact
pressures, contact tractions, contact forces, contact volumes, contact
areas and penetration depth can be derived from a single force chain
network, as shown in Section \ref{subsec:Spherical-particle-particle}.
In particular, we expect traction chain networks to be constant for
spherical particle systems when considering contact laws that scale
linearly with penetration distance as with the LSD.

All DEM simulations in this study were conducted using the LSD model
that includes sliding friction.

\section{\textcolor{black}{Numerical Results\label{sec:Numerical-Results}}}

We first consider force and traction chain networks for spherical
particle systems with and without friction and damping in Section
\ref{subsec:Spherical-particle-particle}. This is followed by a systematic
breakdown of force and traction chain networks for polyhedral particle
systems in Section \ref{subsec:Polyhedral-particle-particle-con}.
Three foundational example problems that isolate and demonstrate the
differences between force chain and traction chain networks are considered
in Section \ref{sec:foundational}. Naturally, these foundational
example problems can serve as unit tests for discrete element codes
and postprocessing scripts. This followed by a septenary particle
stacking that highlights differences between force chain and traction
chain networks for a slightly larger system. Lastly, we conclude with
a 2000 particle system to highlight the complementary nature of force
and traction chain networks. 

\begin{figure}[H]
\begin{centering}
\includegraphics[width=0.8\textwidth]{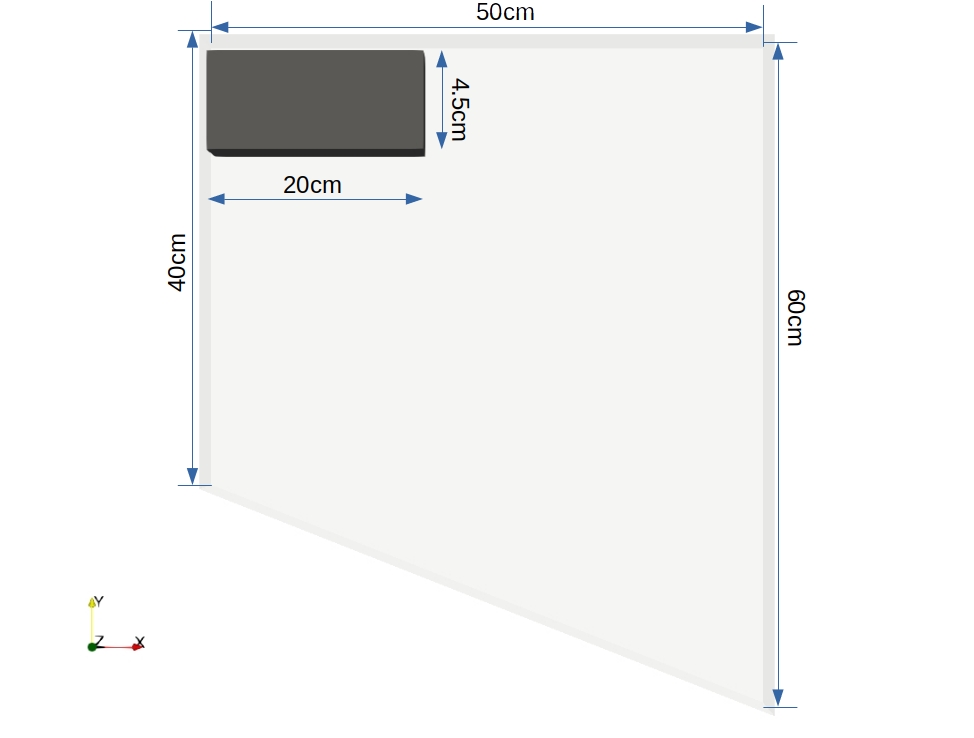}
\par\end{centering}
\caption{The slide box setup with the particle inlet indicated in grey.\label{fig:tiltbox-setup}}
\end{figure}

The geometry of the primary setup is given in planar view as depicted
in Figure \ref{fig:tiltbox-setup}. The out of page dimension of the
container is 5 cm. The particle inlet is indicated in dark grey. The
particles are deposited with a -0.5 m/s vertical velocity. The particle-wall
static and kinetic coefficients of friction are 1.0, the coefficient
of restitution is 0.01. For spherical particle systems, the particle-wall
normal stiffness and tangential stiffnesses are 5000 $N.m^{-1}$ of
2500 $N.m^{-1}$, respectively. For polyhedral particle systems, the
particle-wall normal stiffness and tangential stiffnesses are 5000
$N.m^{-3}$ of 2500 $N.m^{-3}$, respectively. The particle-wall rolling
damping 0.01. 

The particle-particle kinetic coefficients of friction are 0.0 or
1.0 for the frictionless and frictional cases. The particle-particle
coefficient of restitution is 0.01 and 1.0 for the damped and undamped
cases. For spherical particle systems, the particle-wall normal stiffness
and tangential stiffnesses are 5000 $N.m^{-1}$ and 2500 $N.m^{-1}$,
respectively. For polyhedral particle systems, the particle-particle
normal and tangential stiffnesses are 5000 $N.m^{-3}$ and 2500 $N.m^{-3}$,
respectively. The particle-particle rolling damping is 0.001.

In this study, all simulations are subject to a conservative gravitational
field of -9.81 m/s$^{2}$ in the y-direction of indicated reference
frames. Slide box simulations conducted with spherical particles have
diameter 14 mm, cube particles have side lenghts 10 mm having a density
of 2000 $kg/m^{3}$. The spherical and cubic particle systems each
contain around 1150 particles, which is after 1s and 2s of simulation
time, respectively.

All simulations are conducted using the BlazeDEM-GPU framework \citep{Govender2016}.
The BlazeDEM-GPU framework was developed in consortium with the University
of Pretoria, South Africa and the Council for Scientific and Industrial
Research (CSIR), South Africa. The project received significant financial,
experimental and human capital support from IMT Lille Douai, France
enabled upscaling to industrial-scale research using BlazeDEM-GPU
\citep{govender2013development,govender2014Poly,Govender2016,GOVENDER2018}.
Govender is the primary developer of BlazeDEM-GPU, and continually
extending the software's capabilities to expand application domains
\citep{ZHENG2021,ZHENG2021924,GOVENDER2021336}. Development contributions
to BlazeDEM-GPU include the extension of the broad-phase grid-based
spatial partitioning to the boundary-volume hierarchy (BVH) by Lubbe
et al. \citep{lubbe2018,Lubbe_2020}. BlazeDEM-GPU, initially considered
penetration distance to resolve particle-particle contact \citep{govender2013development,govender2014Poly,govender2015Ballmill,Govender2016}.
However, this resulted in instabilities in determining contact directions
\citep{Govender2016,Wilke2017,GOVENDER2018_NONCONVEX}. Wilke et al.
\citep{wilke2016computing} proposed an efficient GPU-based volume
contact algorithm that enabled the extension from convex to non-convex
polyhedra with stable directions. This allowed for stable contact
resolution of convex and non-convex particles in BlazeDEM-GPU using
polyhedral volume overlap \citep{wilke2016computing,GOVENDER20182476}.
Recent, multi-physics extensions include DEM coupling with gradient
corrected SPH by Joubert et al. \citep{joubert2020}. 

In this study, all force chain networks depict the total contact force
magnitude, and all traction networks are computed using the total
contact force magnitude. Additional insights can be gained when decomposing
the total contact forces and tractions into their respective normal
and tangential components.

\subsection{Spherical particle contact\label{subsec:Spherical-particle-particle}}

In the simulations that follow interactions between particles and
the container boundary are non-conservative, i.e. friction and damping
are present using the model parameters outlined in Section \ref{sec:Numerical-Results}.
All simulations represent 1s of simulation time which equates to around
1150 particles having been introduced into the system.

\subsubsection{Conservative contact}

\begin{figure}[H]
\begin{centering}
\includegraphics[width=0.33\textwidth]{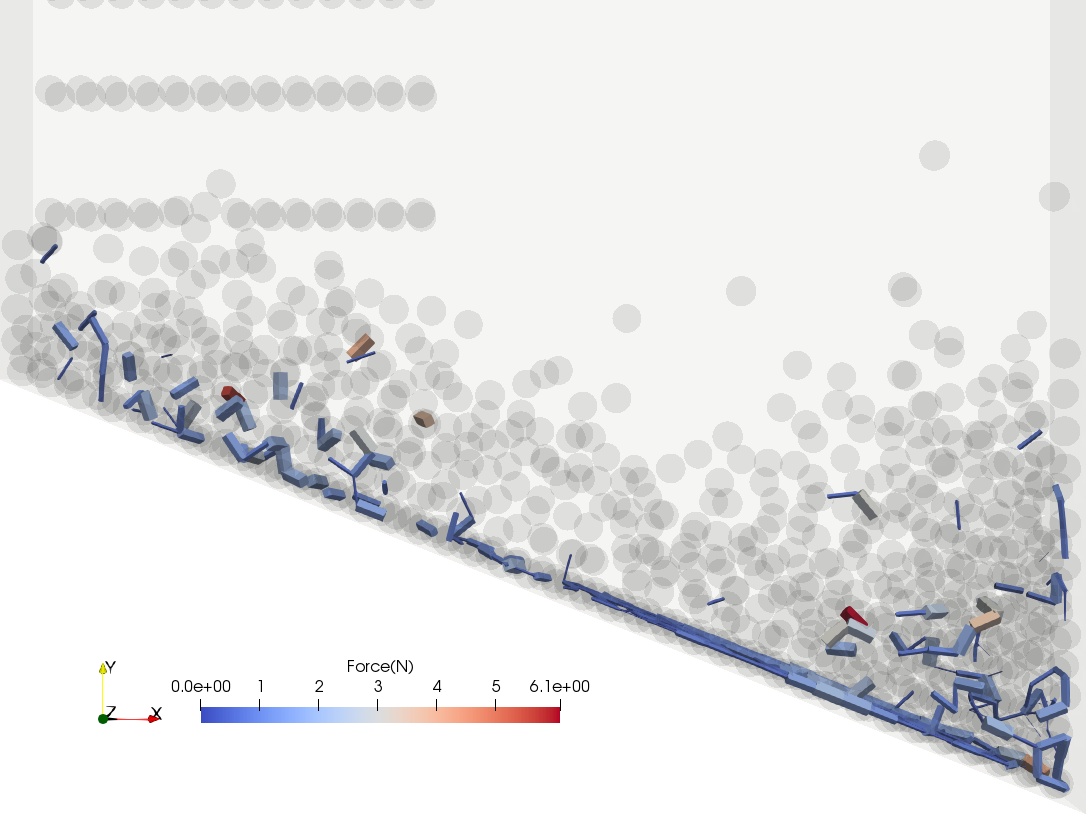}\includegraphics[width=0.33\textwidth]{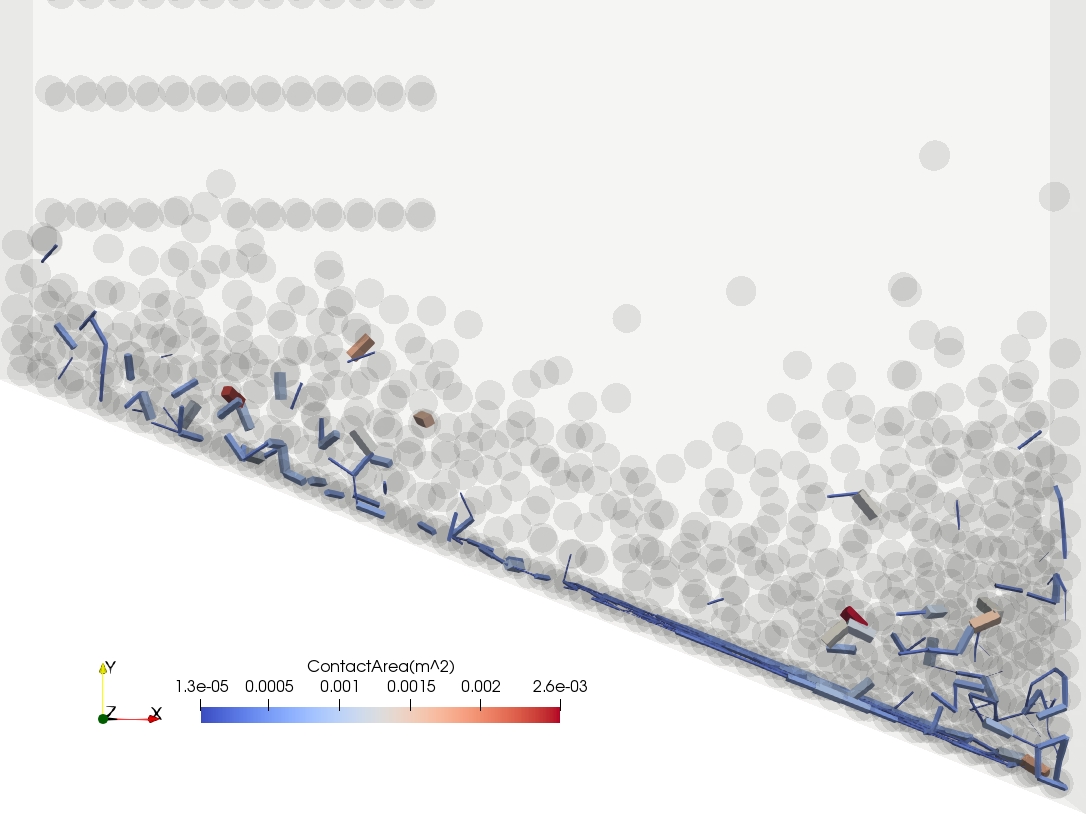}\includegraphics[width=0.33\textwidth]{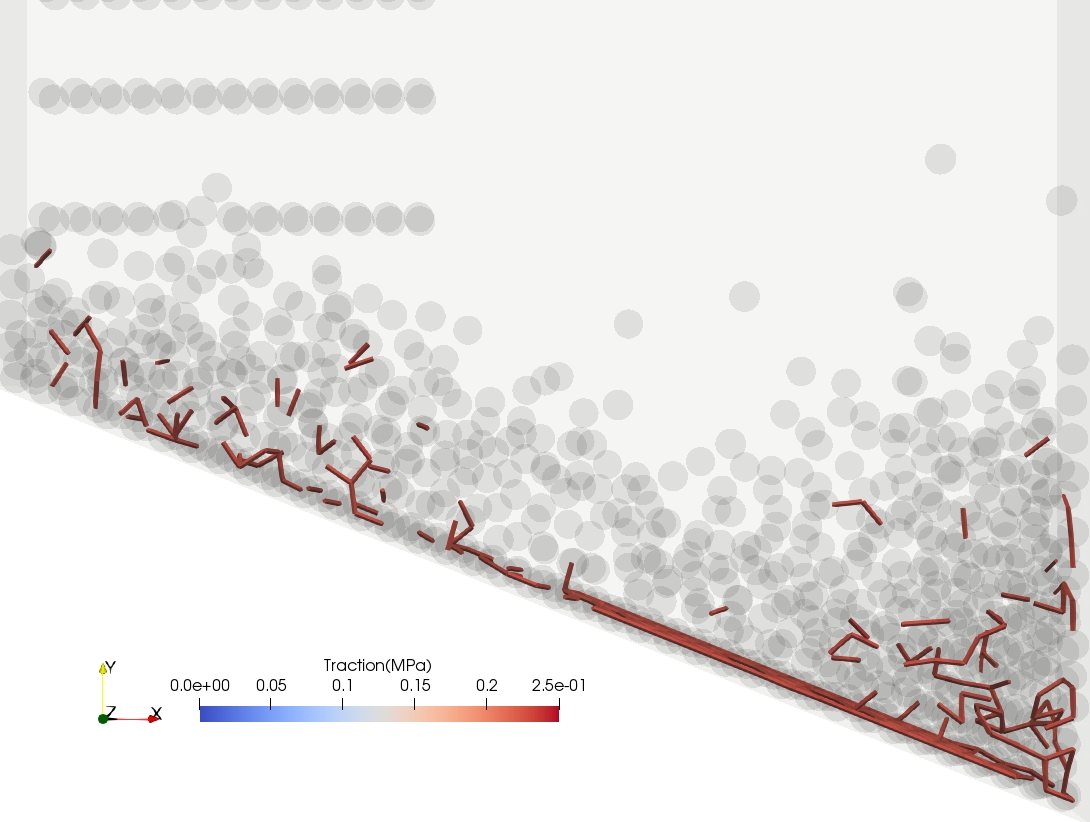}
\par\end{centering}
\caption{Frictionless and undamped: Force, area and traction chain networks
for a spherical particle system with conservative particle-particle
(frictionless and undamped) interactions.}

\label{fig:forcechain-1}
\end{figure}

Consider three information networks, force, contact area and traction
depicted in Figure \ref{fig:forcechain-1}, for a conservative spherical
particle system, i.e. for frictionless and undamped particle-particle
contact. The force chain and area chain networks are merely scaled
versions of the same information. The traction chain network is constant
and adds no information beyond contributing the scalar value $\approx0.23$
MPa, which is the result of using a linear spring contact model. The
$0.23$ MPa equates to a normal spring stiffness of 5000 $N.m^{-1}$.

\subsubsection{Damped frictionless contact}

\begin{figure}[H]
\begin{centering}
\includegraphics[width=0.33\textwidth]{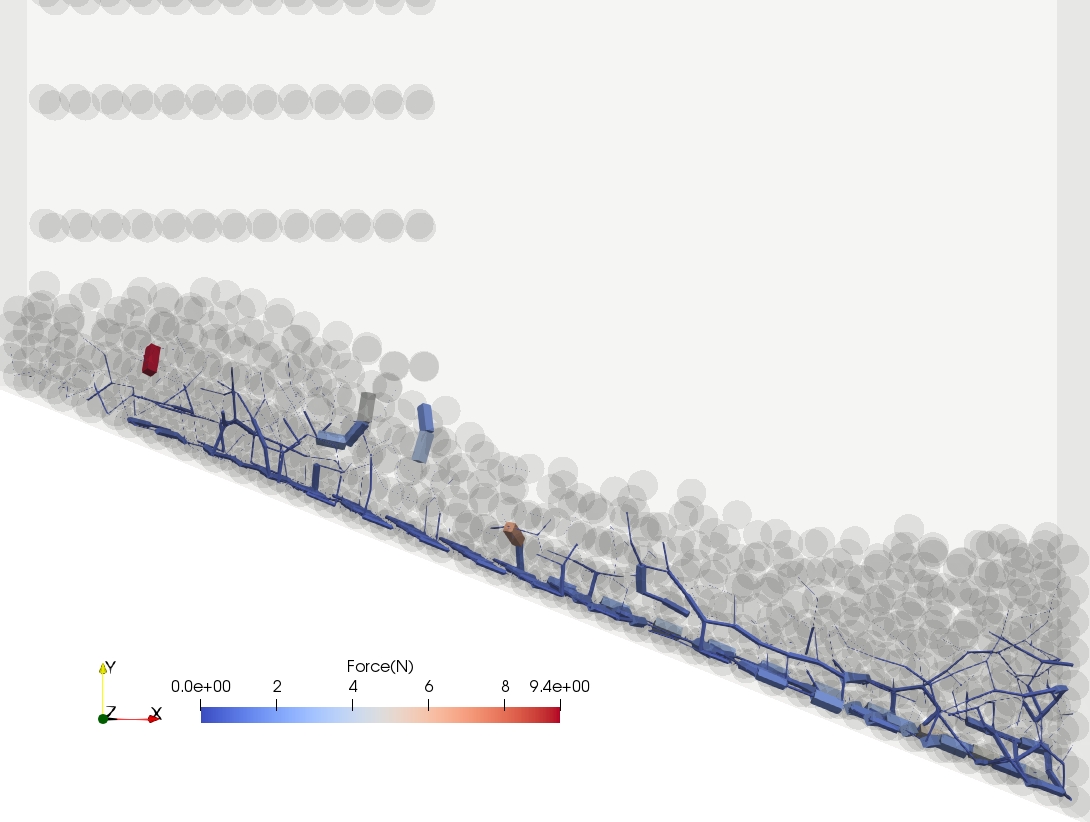}\includegraphics[width=0.33\textwidth]{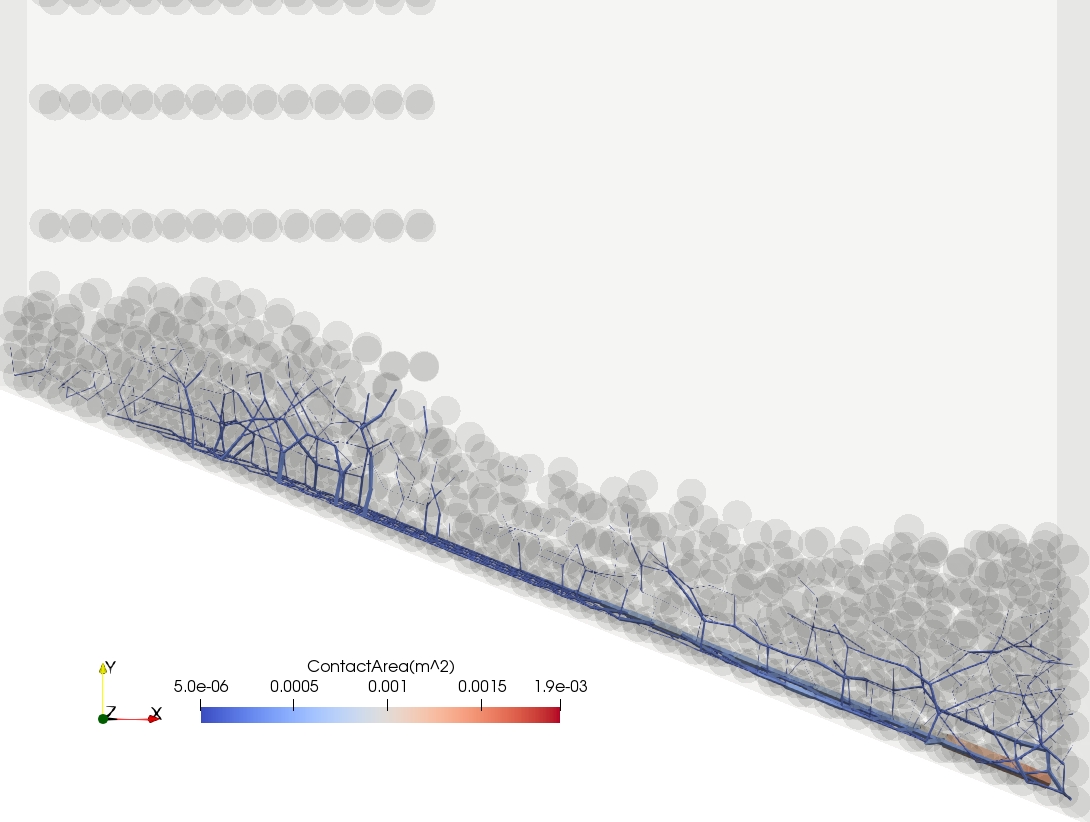}\includegraphics[width=0.33\textwidth]{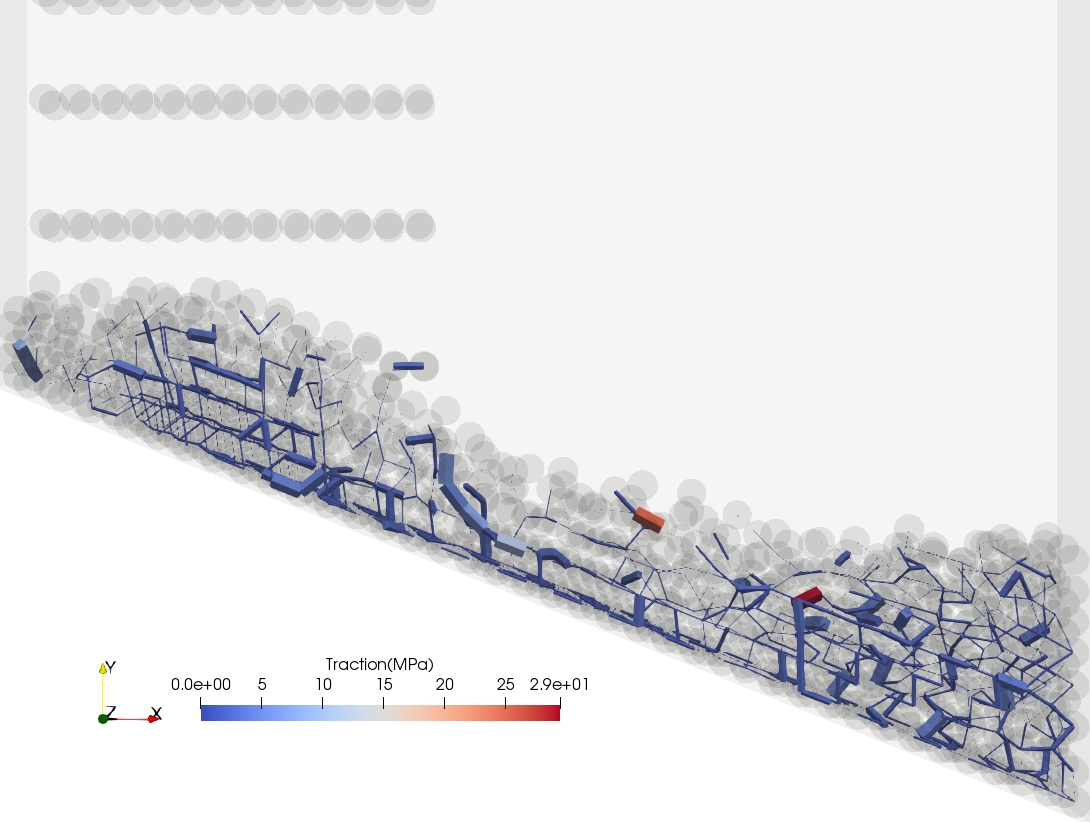}
\par\end{centering}
\caption{Frictionless and damped: Force, area and traction chain networks for
a spherical particle system with non-conservative (damped) particle-particle
interactions.}

\label{fig:forcechain-2}
\end{figure}

Next, consider three information networks, force, contact area and
traction depicted in Figure \ref{fig:forcechain-2}, for a frictionless
but damped particle-particle contact. Here, particle damping contributes
additional forces that the particles experience, in addition to the
linear spring. As a result, dynamic contact result in distinct tractions,
whereas static contact is representative of a homogenous traction
network. Overall, the force, contact area and traction chain networks
are distinct. Differences in the traction network are now only the
result of damping forces that are non-zero in dynamic contact between
particles.

\medskip{}

\subsubsection{Undamped frictional contact}

\begin{figure}[H]
\begin{centering}
\includegraphics[width=0.33\textwidth]{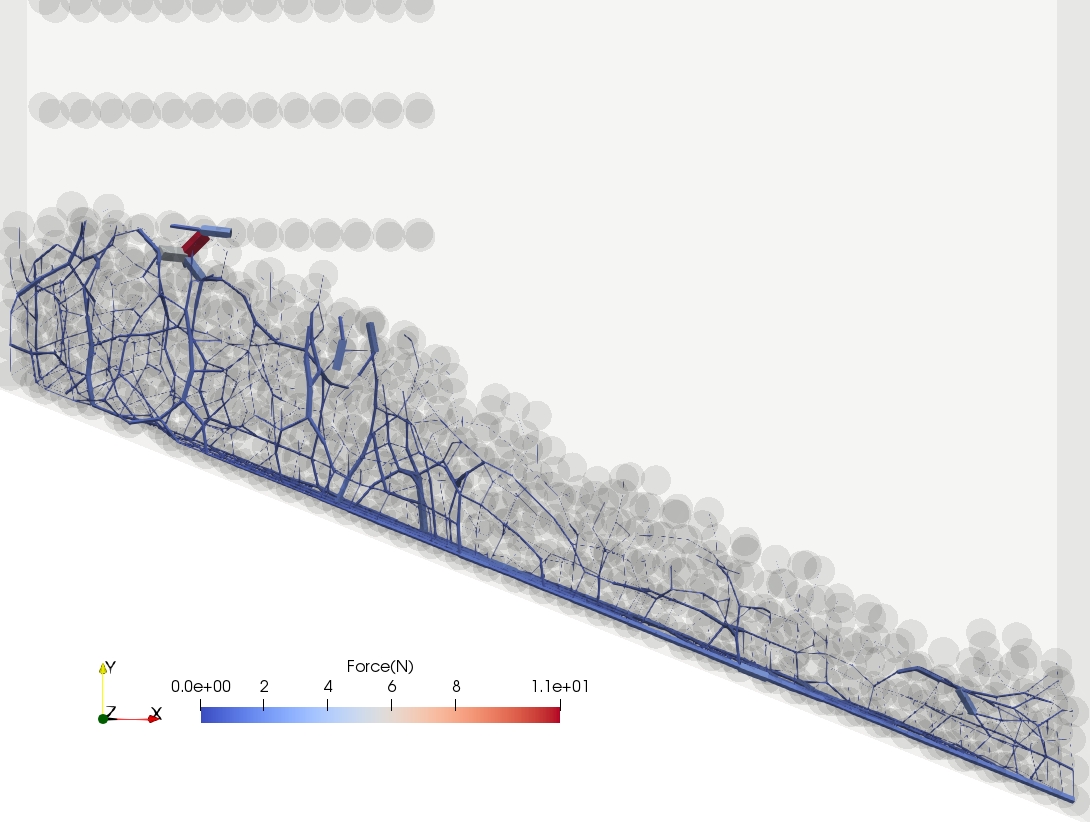}\includegraphics[width=0.33\textwidth]{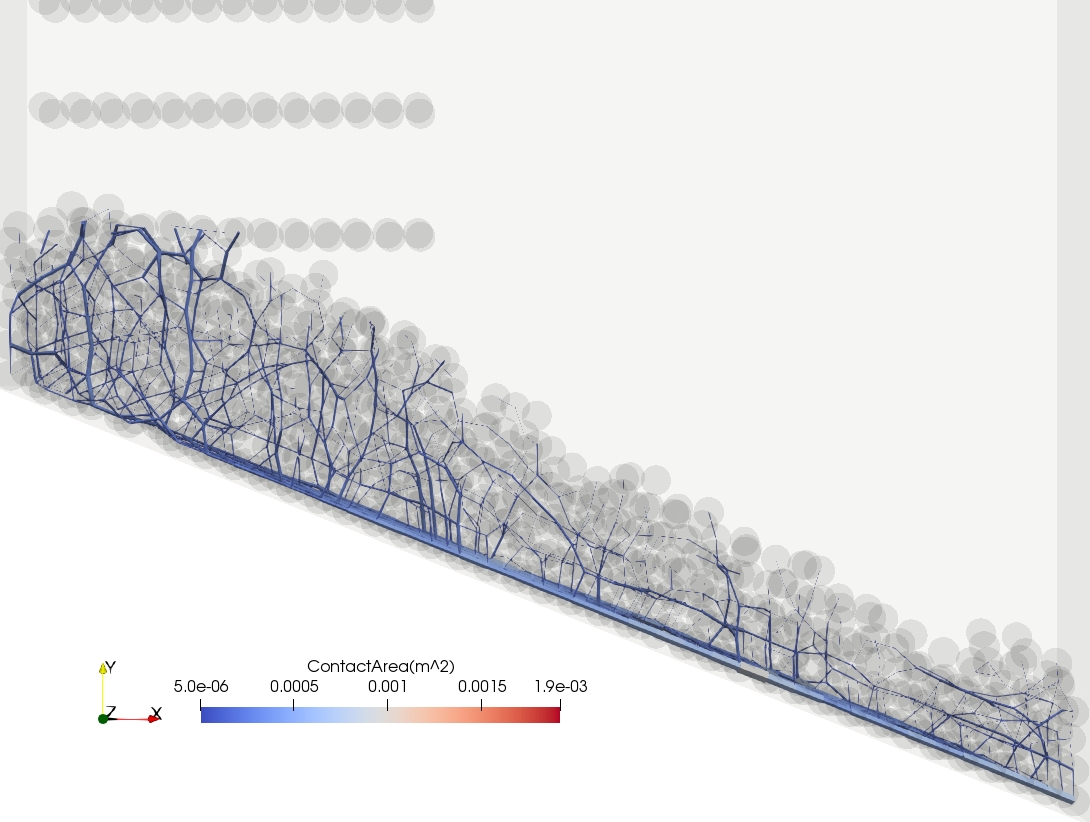}\includegraphics[width=0.33\textwidth]{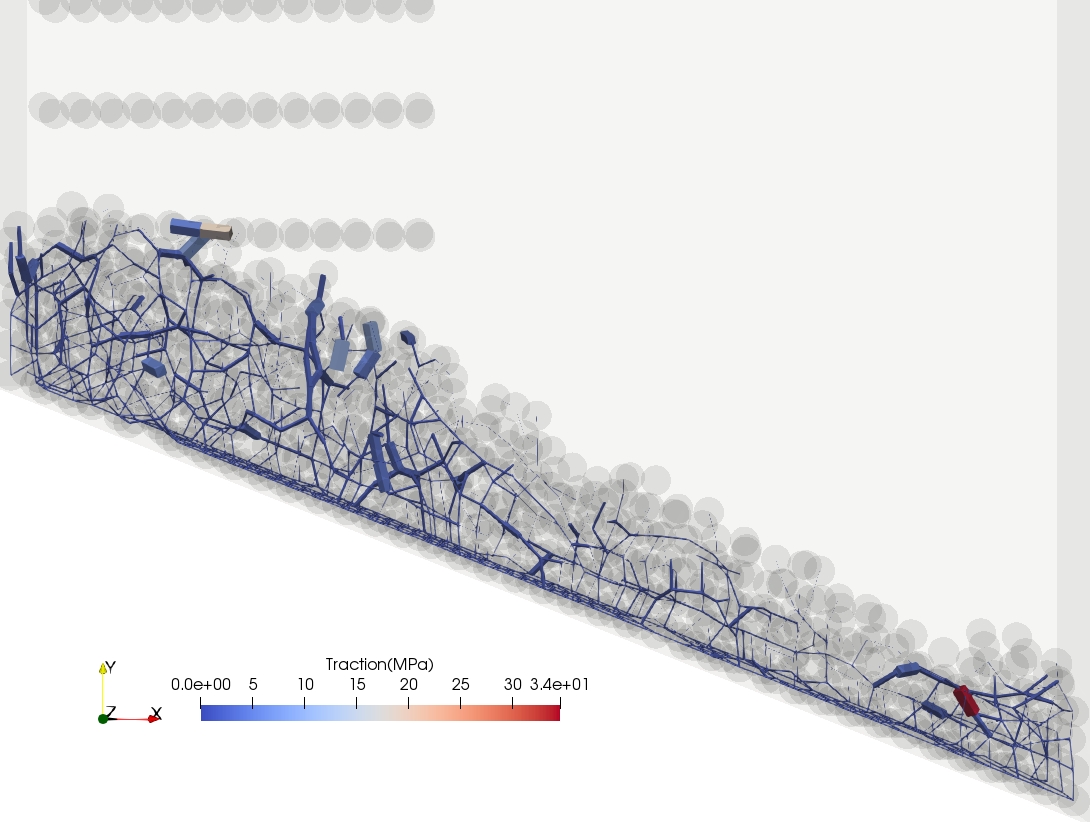}
\par\end{centering}
\caption{Frictional and undamped: Force, area and traction chain networks for
a spherical particle system with non-conservative (frictional) particle-particle
interactions.}

\label{fig:forcechain-3}
\end{figure}

Consider now the three information networks, force, contact area and
traction depicted in Figure \ref{fig:forcechain-3}, for undamped
but frictional particle-particle contact. Here, particle-particle
friction contributes additional forces that the particles experience,
in addition to the linear spring. As a result, static and dynamic
contact result in distinct tractions. Overall, the force, contact
area and traction chain networks are distinct. Differences in the
traction network are now only the result of frictional forces that
are non-zero in both static and dynamic contact between particles.

\subsubsection{Damped frictional contact}

\begin{figure}[H]
\begin{centering}
\includegraphics[width=0.33\textwidth]{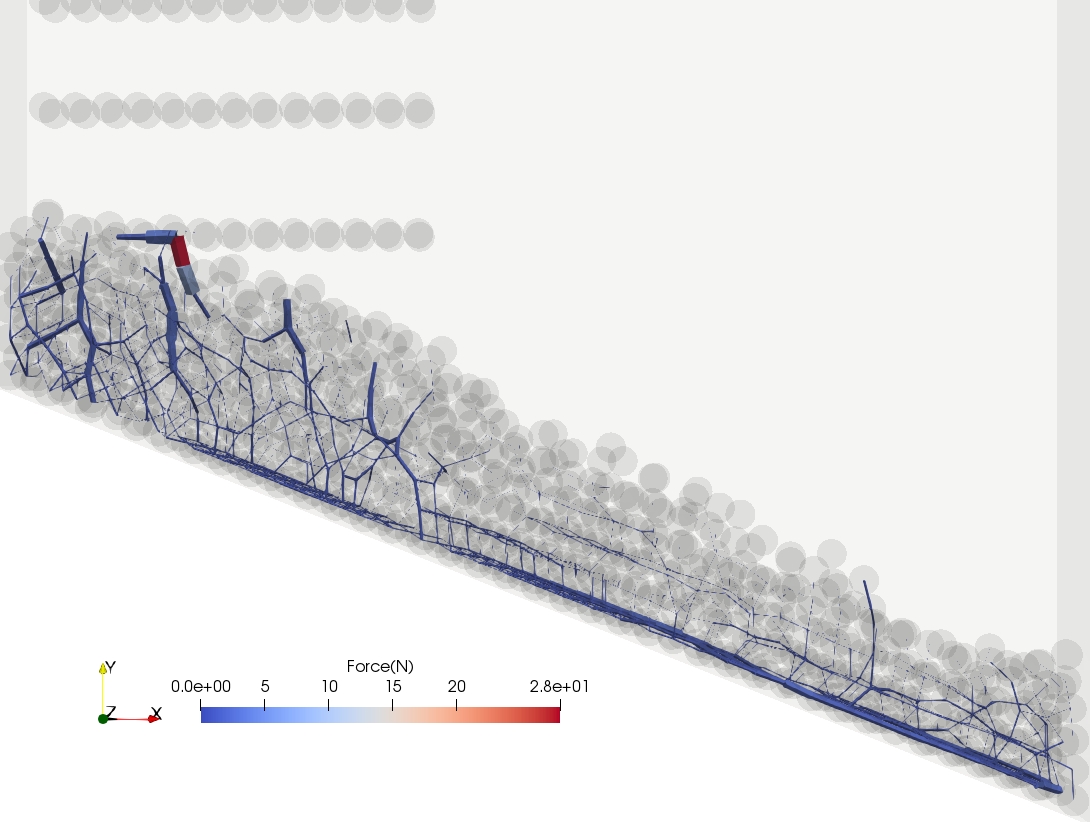}\includegraphics[width=0.33\textwidth]{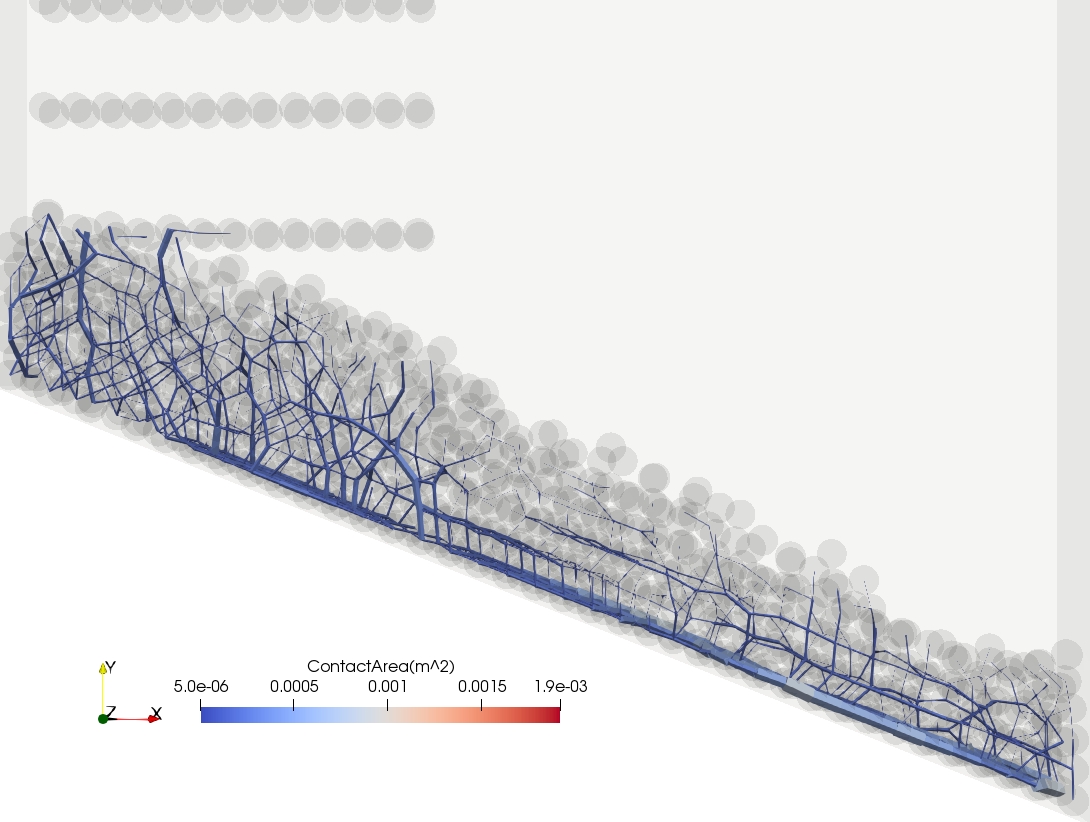}\includegraphics[width=0.33\textwidth]{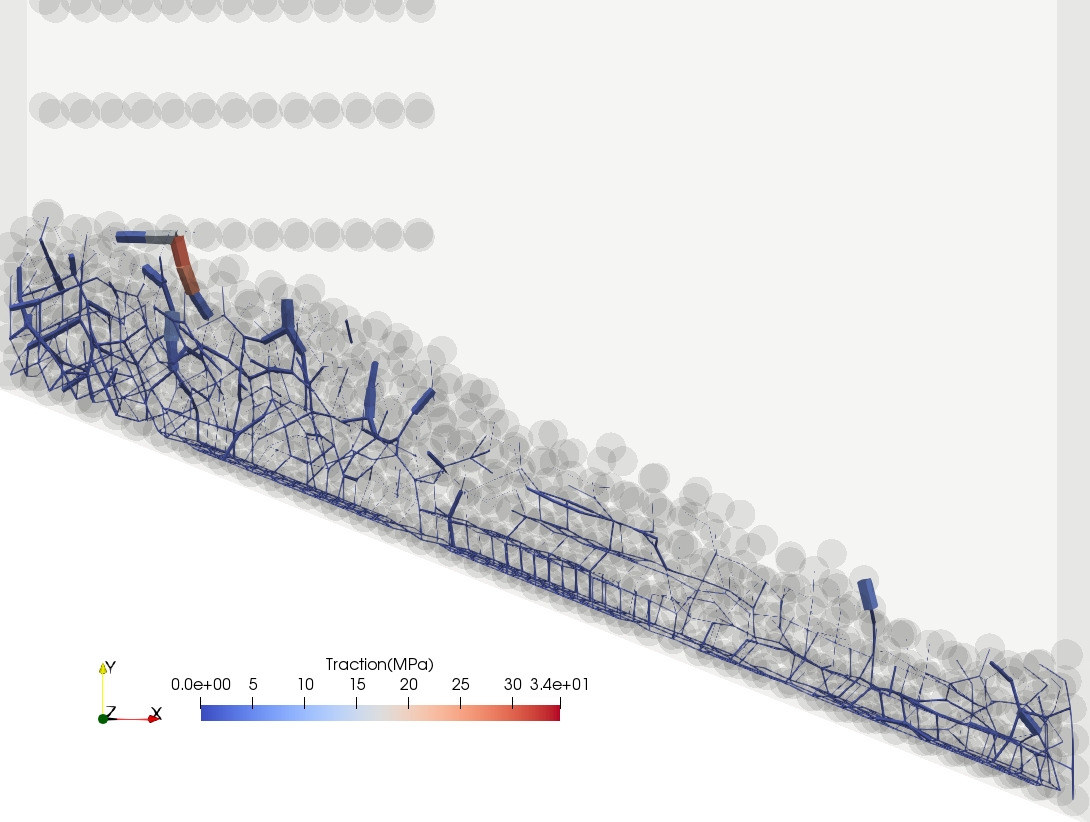}
\par\end{centering}
\caption{Frictional and damped: Force, area and traction chain networks for
a spherical particle system with non-conservative (frictional and
damped) particle-particle interactions.}

\label{fig:forcechain-4}
\end{figure}

Consider now the three information networks, force, contact area and
traction depicted in Figure \ref{fig:forcechain-4}, for damped and
frictional particle-particle contact. Here, damping and particle-particle
friction contributes additional forces that the particles experience,
in addition to the linear spring. As a result, static and dynamic
contact result in distinct tractions. Overall, the force, contact
area and traction chain networks are distinct. Differences in the
traction network are due to both frictional and damping forces that
are non-zero in both static and dynamic contact between particles.

\subsection{Polyhedral particle contact\label{subsec:Polyhedral-particle-particle-con}}

\subsubsection{Binary particle system\label{sec:foundational}}

\begin{figure}[H]
\begin{centering}
\includegraphics[scale=0.25]{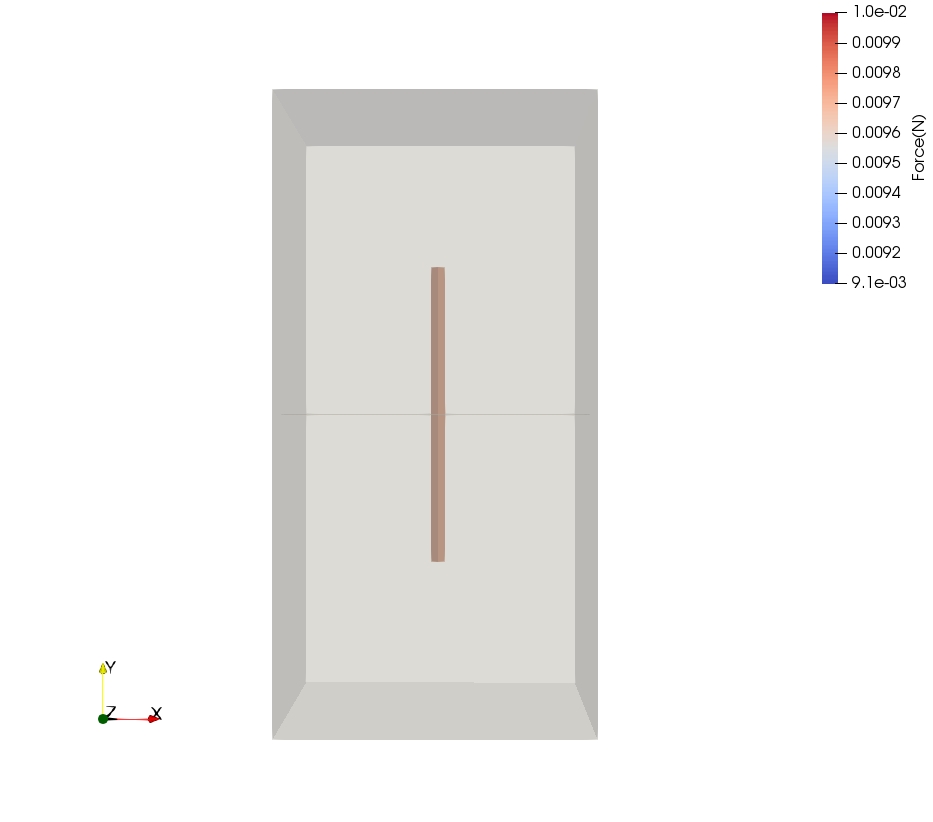}\includegraphics[scale=0.25]{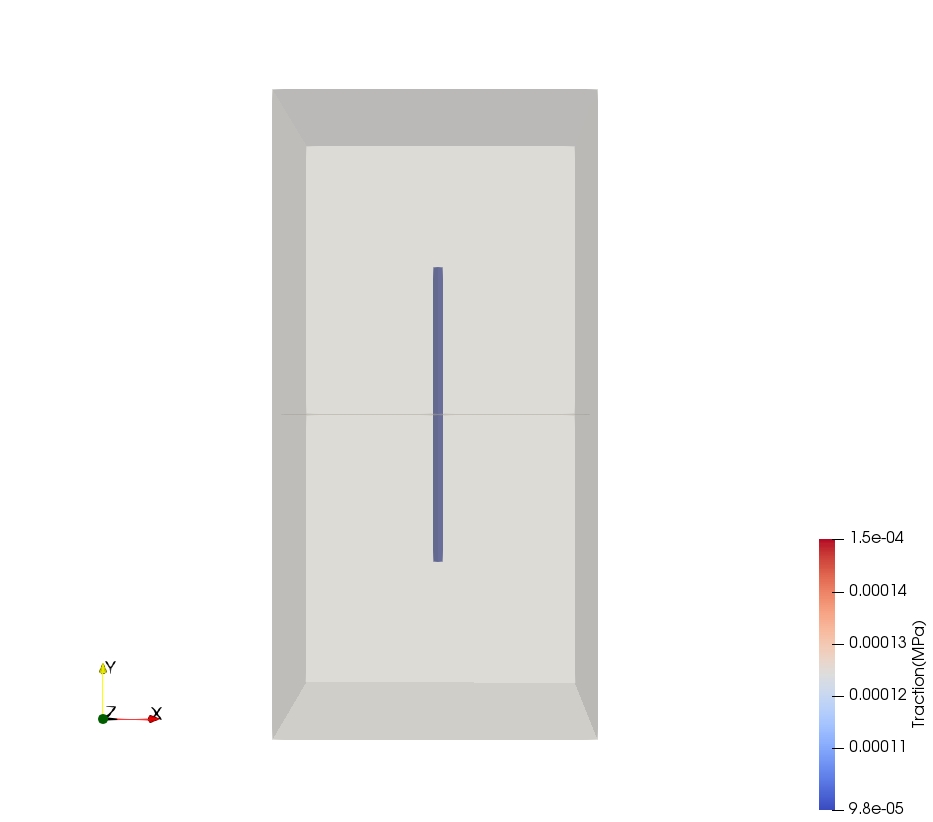}
\par\end{centering}
\caption{Vertical stack: Force chain network (left) and traction chain network
(right) for a binary particle system.\label{fig:cubes_vs}}
\end{figure}

\begin{figure}[H]
\begin{centering}
\includegraphics[scale=0.25]{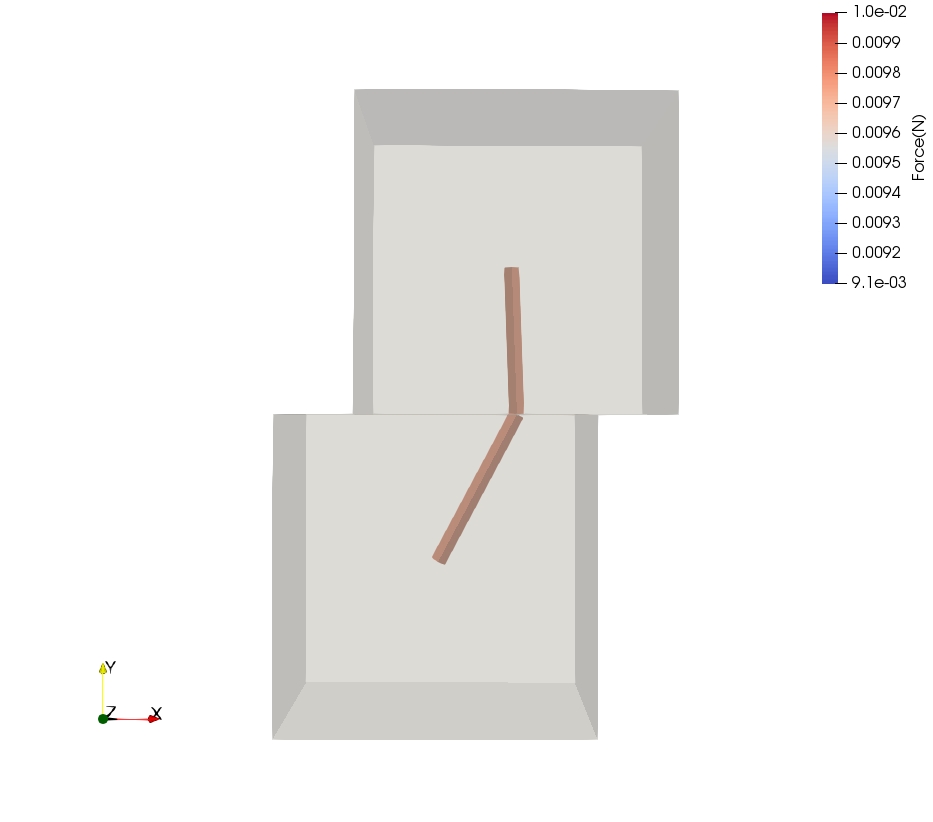}\includegraphics[scale=0.25]{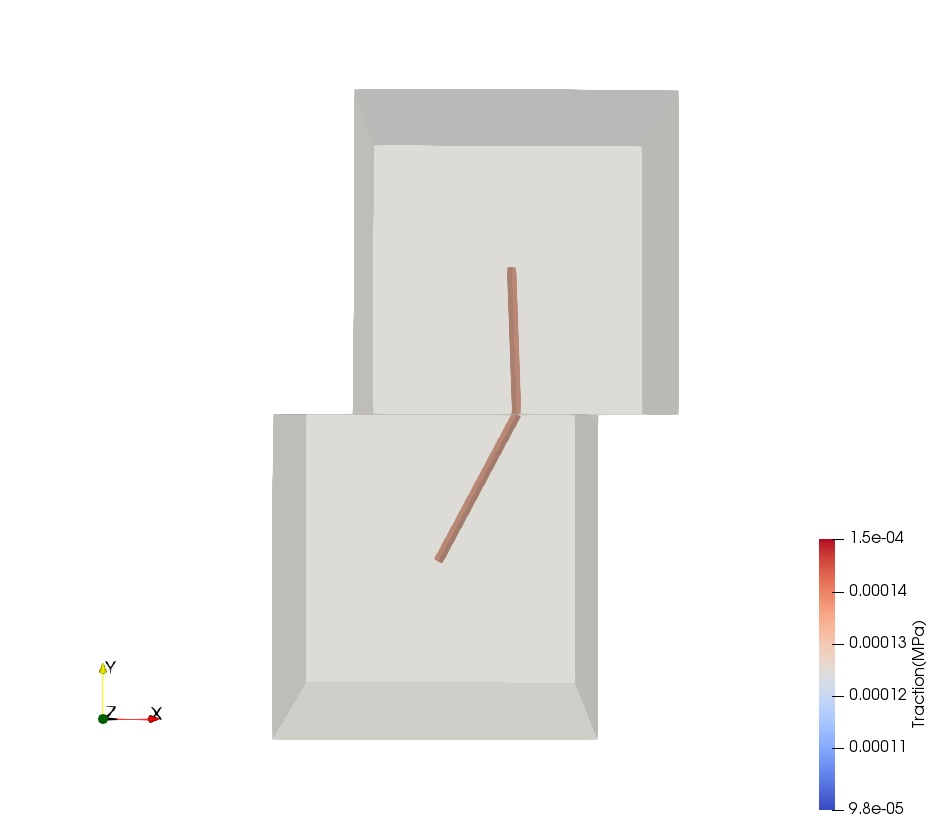}
\par\end{centering}
\caption{Overhang stack: Force chain network (left) and traction chain network
(right) for a binary particle system.\label{fig:cube_os}}
\end{figure}

\begin{figure}[H]
\begin{centering}
\includegraphics[scale=0.25]{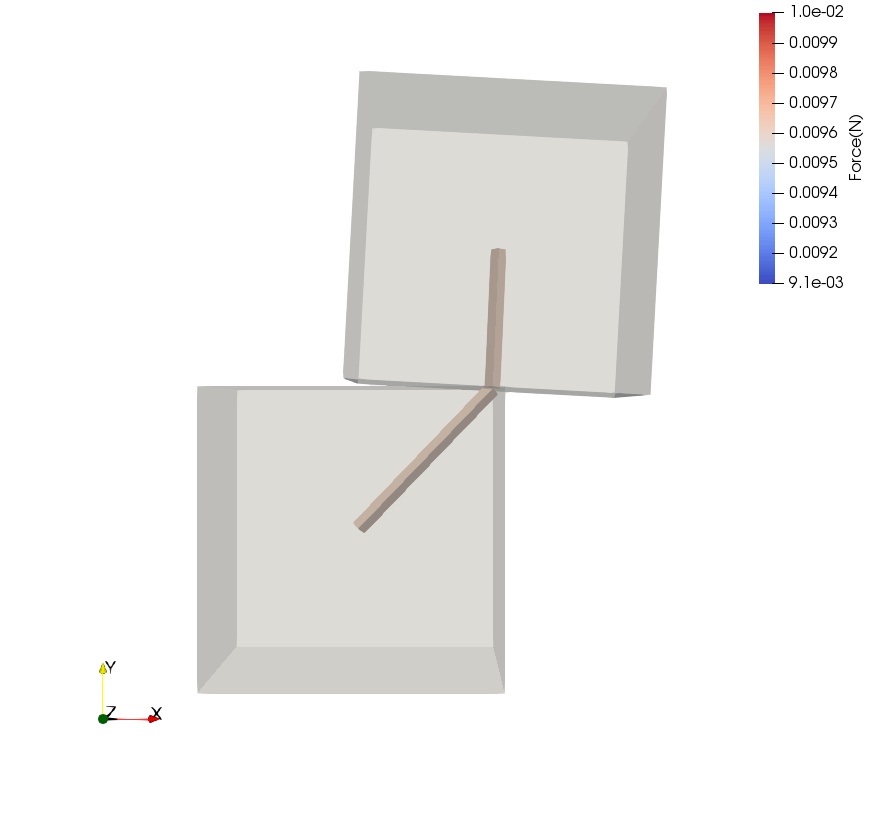}\includegraphics[scale=0.25]{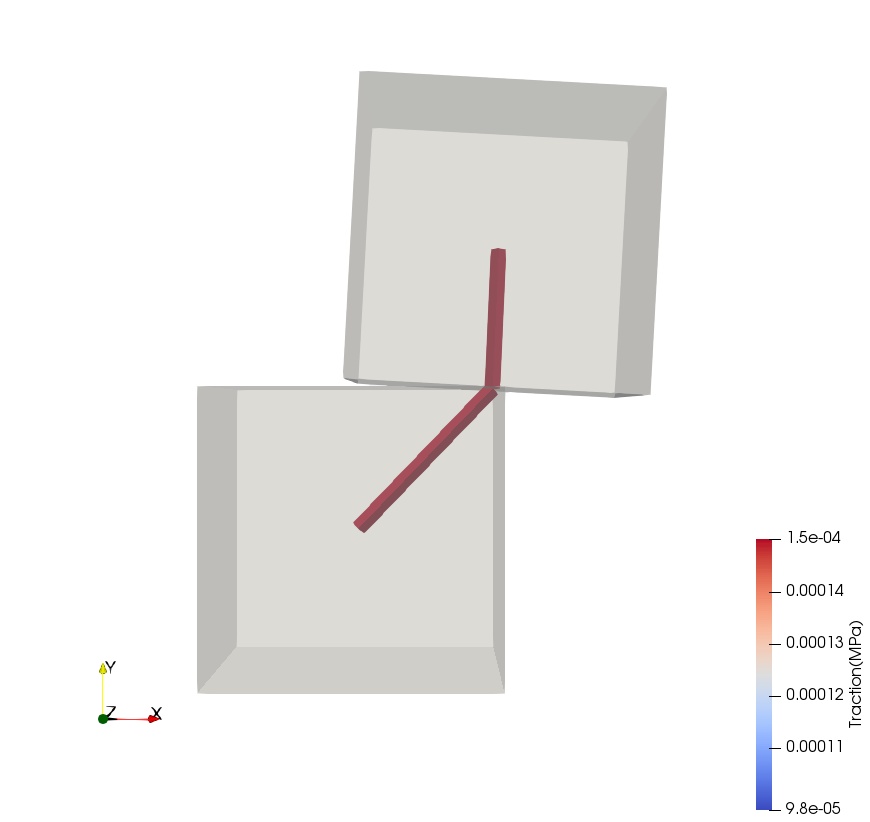}
\par\end{centering}
\caption{Topple stack: Force chain network (left) and traction chain network
(right) for a binary particle system.\label{fig:cube_ts}}
\end{figure}

The cuboids and reference frames are shown in Figures \ref{fig:cubes_vs}-\ref{fig:cube_ts}.
Each has a side length of 1 cm, a density of 1 g/cm$^{3}$ subject
to a conservative gravitational field of -9.81 m/s$^{2}$ in the y-direction.
Three stacking scenarios are considered, namely, a vertical stack
(Figure \ref{fig:cubes_vs}), an overhang stack (Figure \ref{fig:cube_os})
and topple stack (Figure \ref{fig:cube_ts}). The simulation were
conducted with a time step of $1\times10^{-6}s$.

The particles are perfectly aligned for the vertical stack, while
for the overhang stack, the top cube is shifted by 0.25cm. For the
topple stack, the top cube is shifted by 0.5cm.

As a result, the quasi-static vertical reaction force magnitude is
0.00981N in all three scenarios. In turn, for the vertical stack,
the average traction magnitude is 98.1 Pa (9.81$\times10^{-5}$MPa).
For the overhang scenario, the average traction magnitude is 130.8
Pa (1.31$\times10^{-4}$MPa). It is above 196.2 Pa (1.96$\times10^{-4}$MPa)
for the topple scenario as it varies with the resolved contact area,
which in turn depends on the exact particle orientation and contact
stiffness.

\subsubsection{Septenary particle system\label{sec:cube7}}

\begin{figure}[H]
\begin{centering}
\includegraphics[scale=0.2]{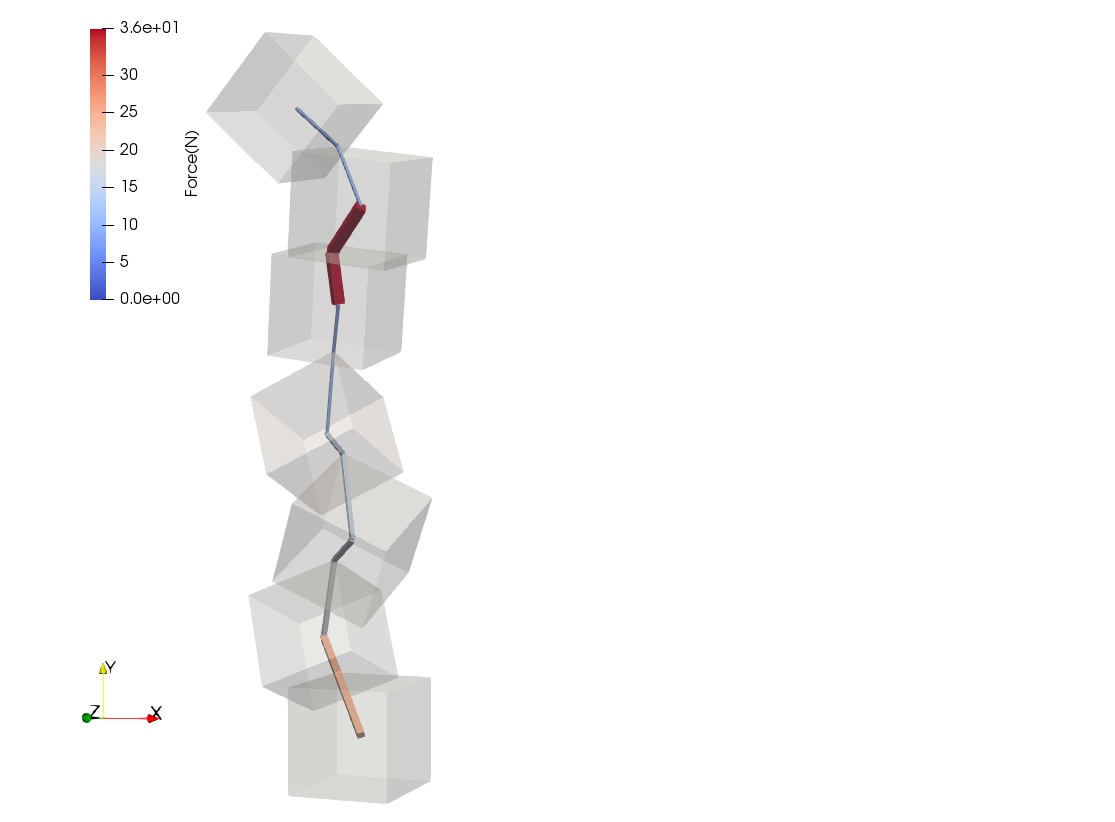}\includegraphics[scale=0.2]{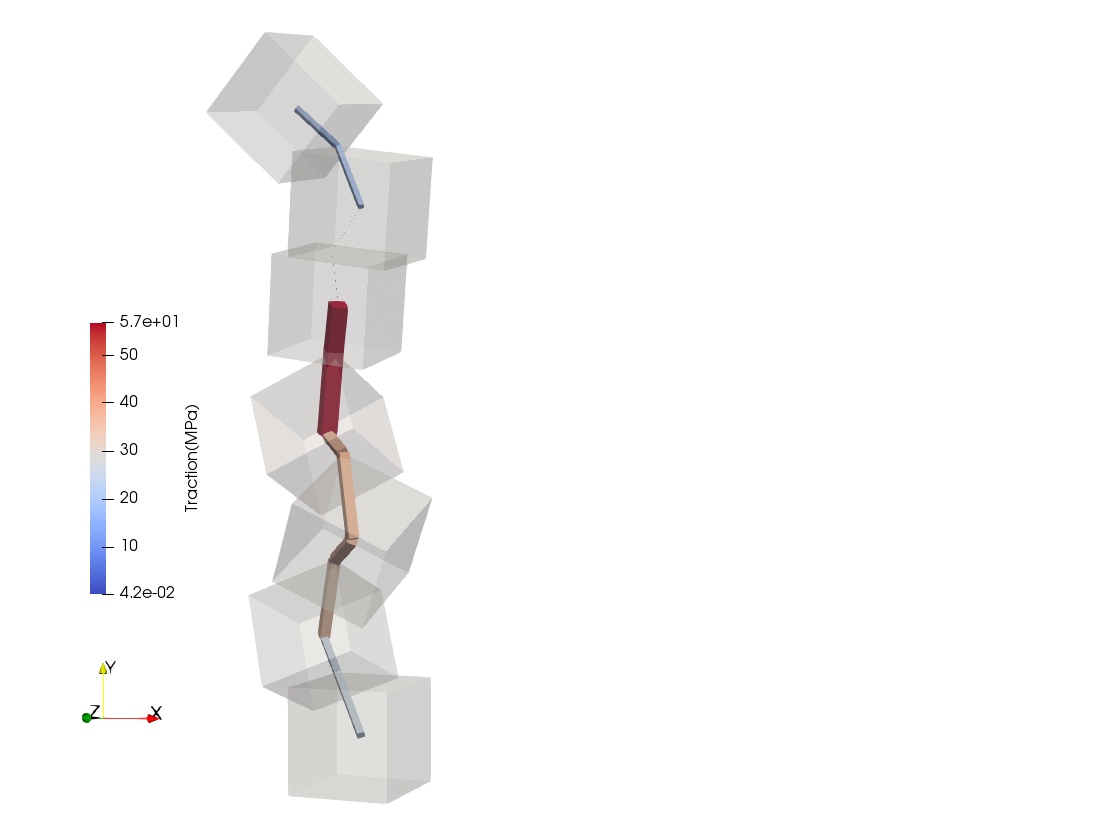}
\par\end{centering}
\caption{Topple stack: Force chain network (left) and traction chain network
(right) for a septenary particle system.\label{fig:7cube}}
\end{figure}

Next, we consider seven cubic aluminium particles weighing 3.375 N
each, i,.e. a side length 50mm loaded into a 100mm by 100mm container
from a height of 450mm. The particles are deposited with a -1 m/s
vertical velocity subject to a conservative gravitational field of
-9.81 m/s$^{2}$ in the y-direction on the indicated reference frame
in Figure \ref{fig:7cube}. 

The particle-wall static and kinetic coefficients of friction are
0.6, the coefficient of restitution is 0.4. The particle-wall normal
stiffness and tangential stiffnesses are 5000 $N.m^{-3}$ and 2500
$N.m^{-3}$, respectively. The particle-particle kinetic coefficients
of friction are 0.0, the coefficient of restitution is 0.1. The normal
stiffness is 5000 $N.m^{-3}$ and the tangential stiffness is 2500
$N.m^{-3}$. The simulation were conducted with a time step of $5\times10^{-5}s$.

The largest force in the force chain network is distinct from the
largest traction in the traction chain network, which demonstrates
the complementary nature of the information contained in the two networks.
The largest forces are clearly associated with large contact areas,
which result in the lowest contact pressure as quantified by the traction
network. Consequently, the force chain network only gives partial
insight to inter-particle contact, which the traction chain network
complements. Similarly, the traction chain network when viewed in
isolation only partially explains inter-particle contact. This highlights
the importance of considering both force chain and traction chain
networks when interpreting non-spherical granular systems.

\subsubsection{Large force and traction networks\label{sec:shape}}

\begin{figure}[H]
\begin{centering}
\includegraphics[width=0.33\textwidth]{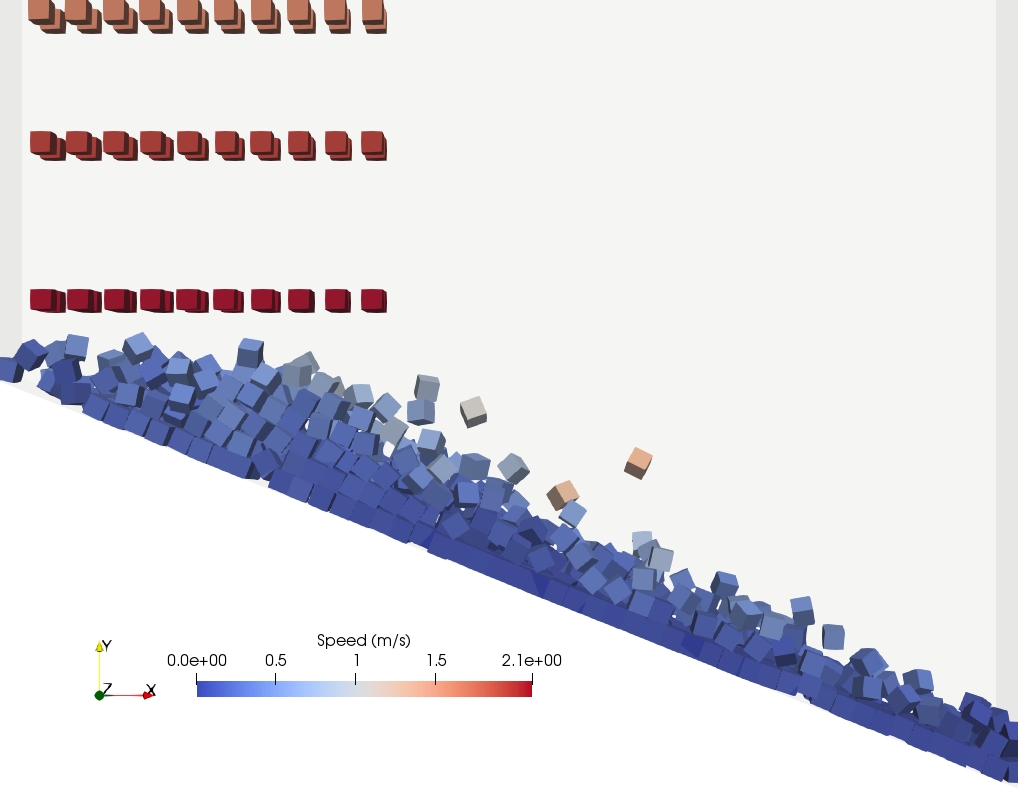}\includegraphics[width=0.33\textwidth]{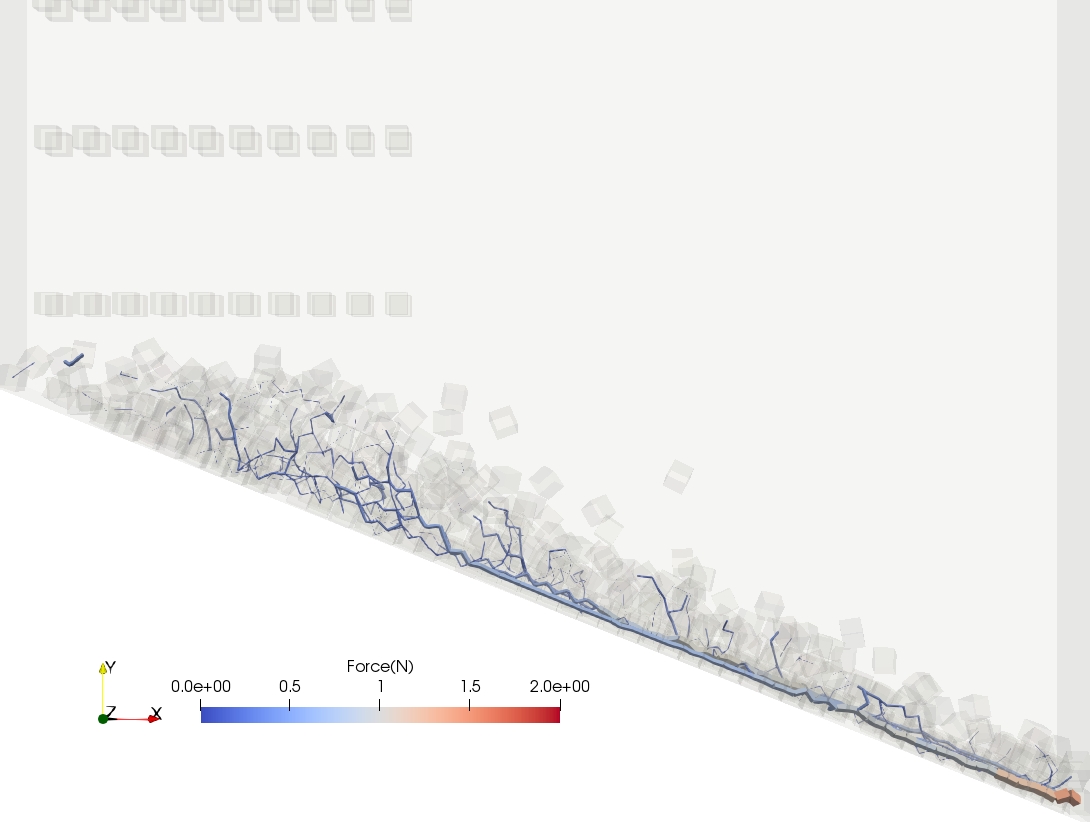}\includegraphics[width=0.33\textwidth]{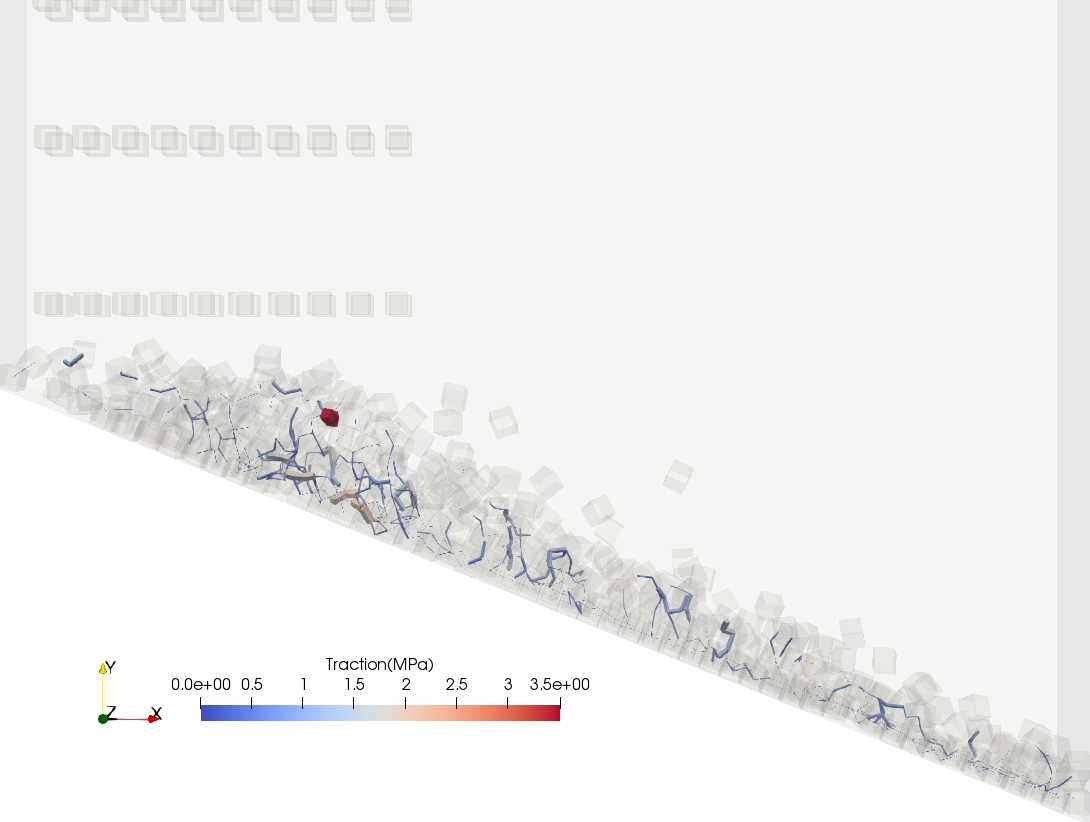}
\par\end{centering}
\begin{centering}
\includegraphics[width=0.33\textwidth]{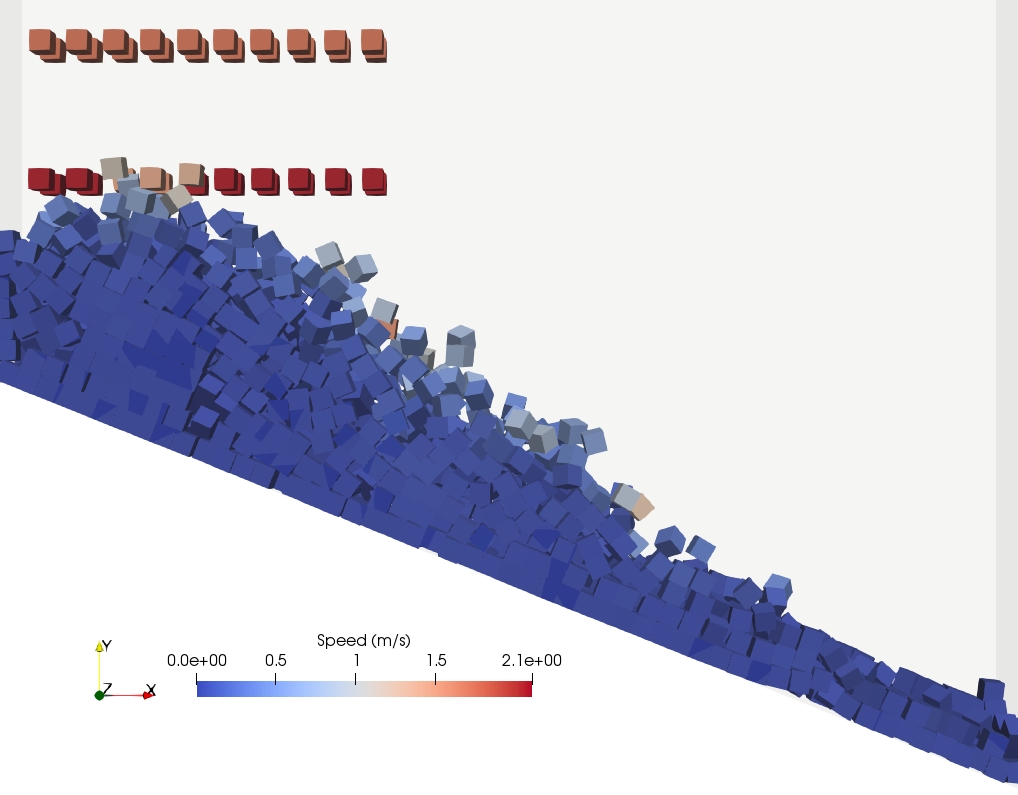}\includegraphics[width=0.33\textwidth]{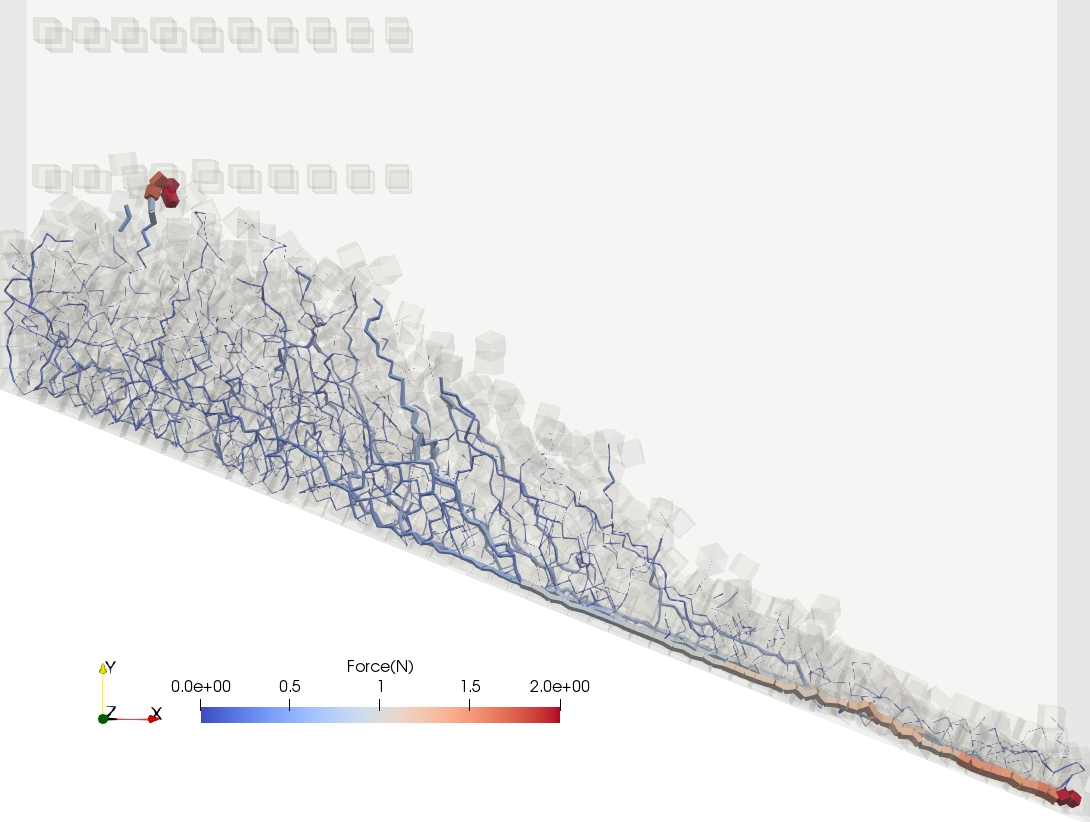}\includegraphics[width=0.33\textwidth]{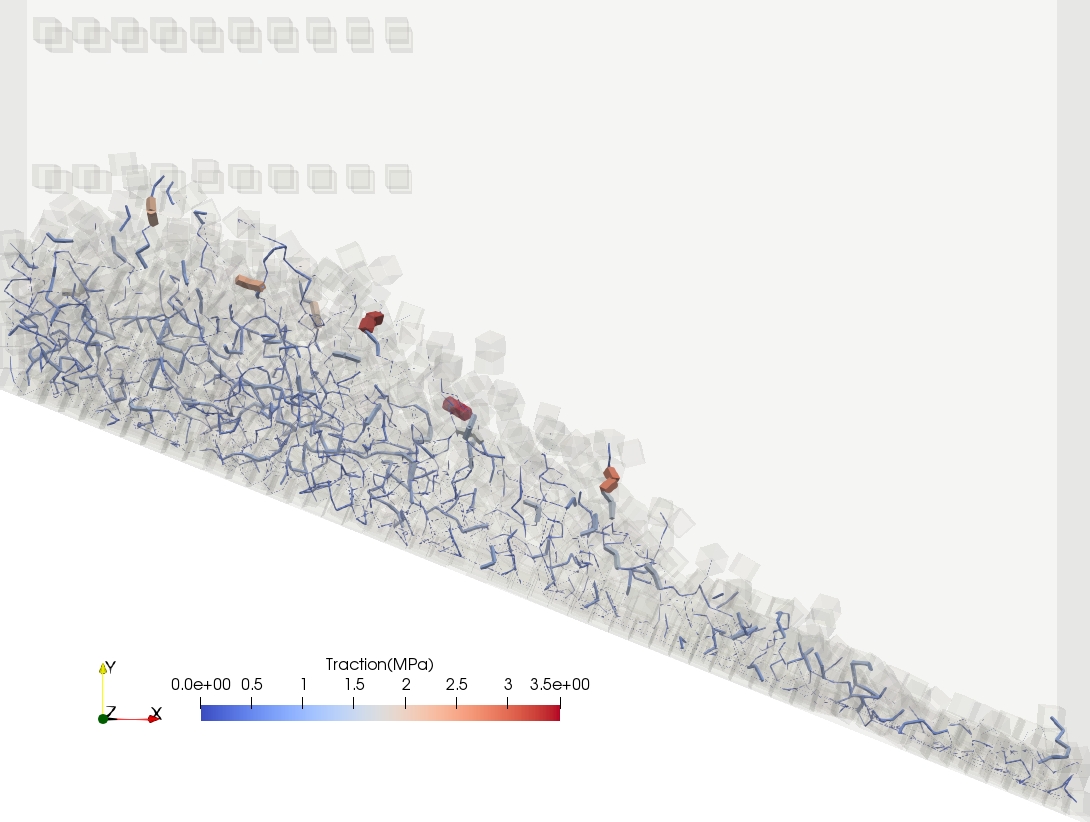}
\par\end{centering}
\begin{centering}
\includegraphics[width=0.33\textwidth]{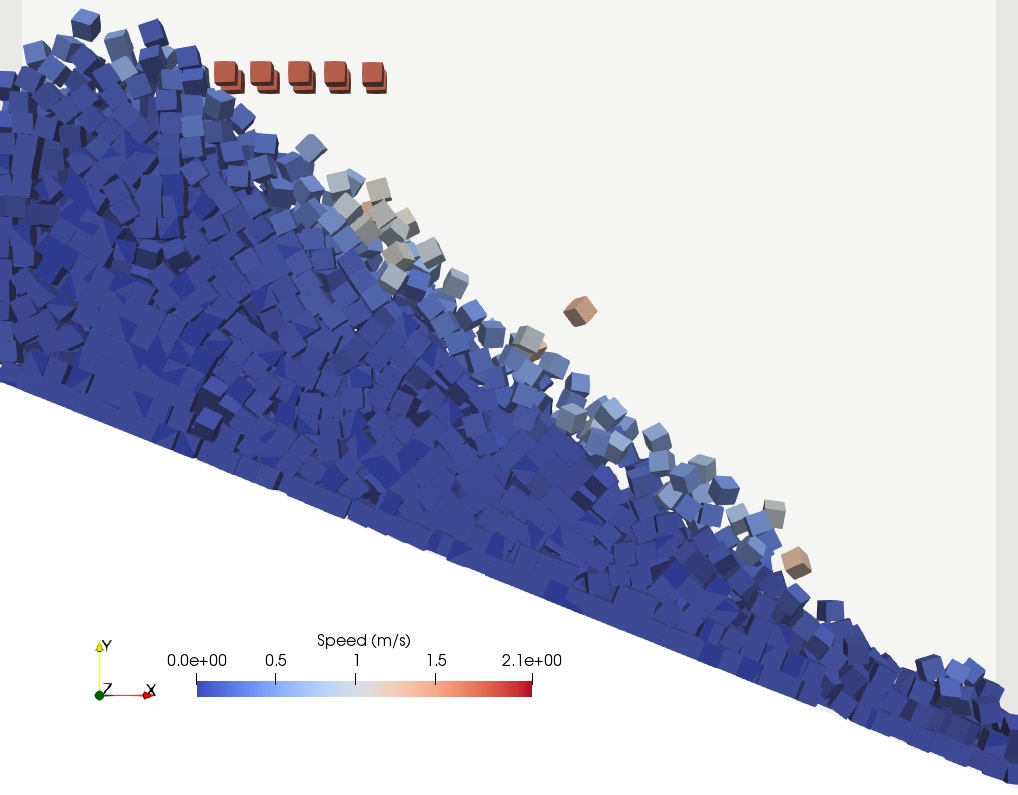}\includegraphics[width=0.33\textwidth]{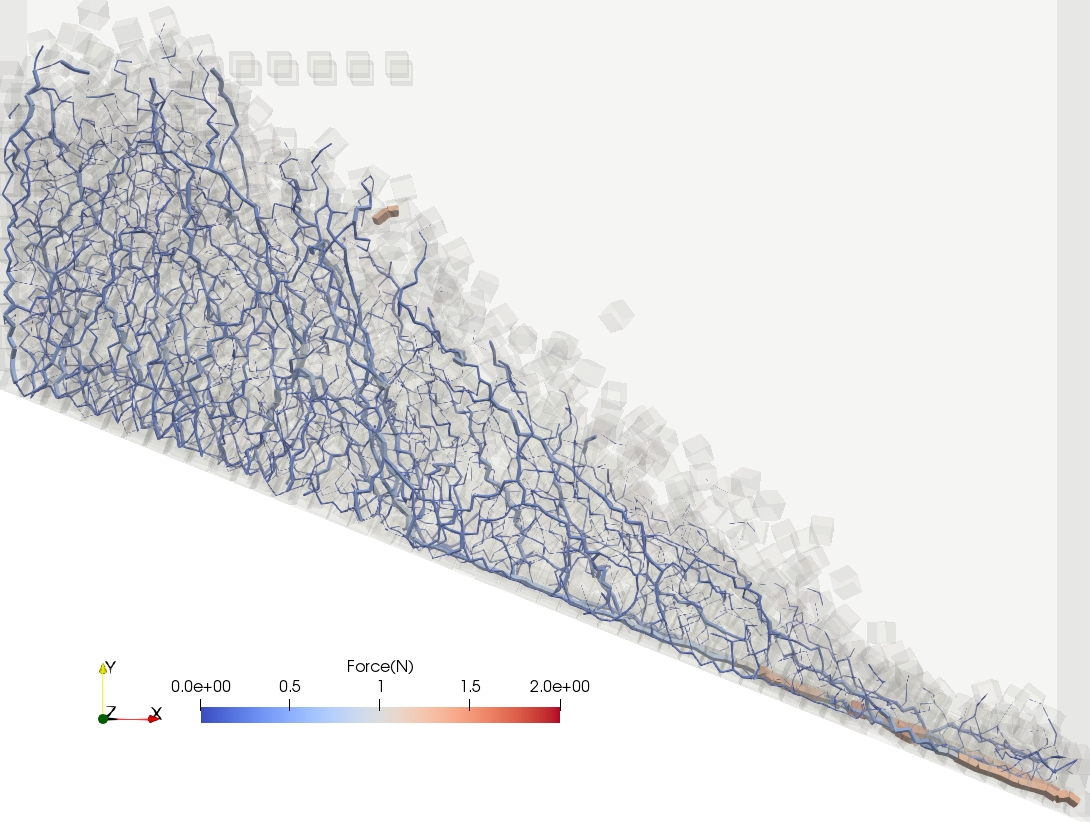}\includegraphics[width=0.33\textwidth]{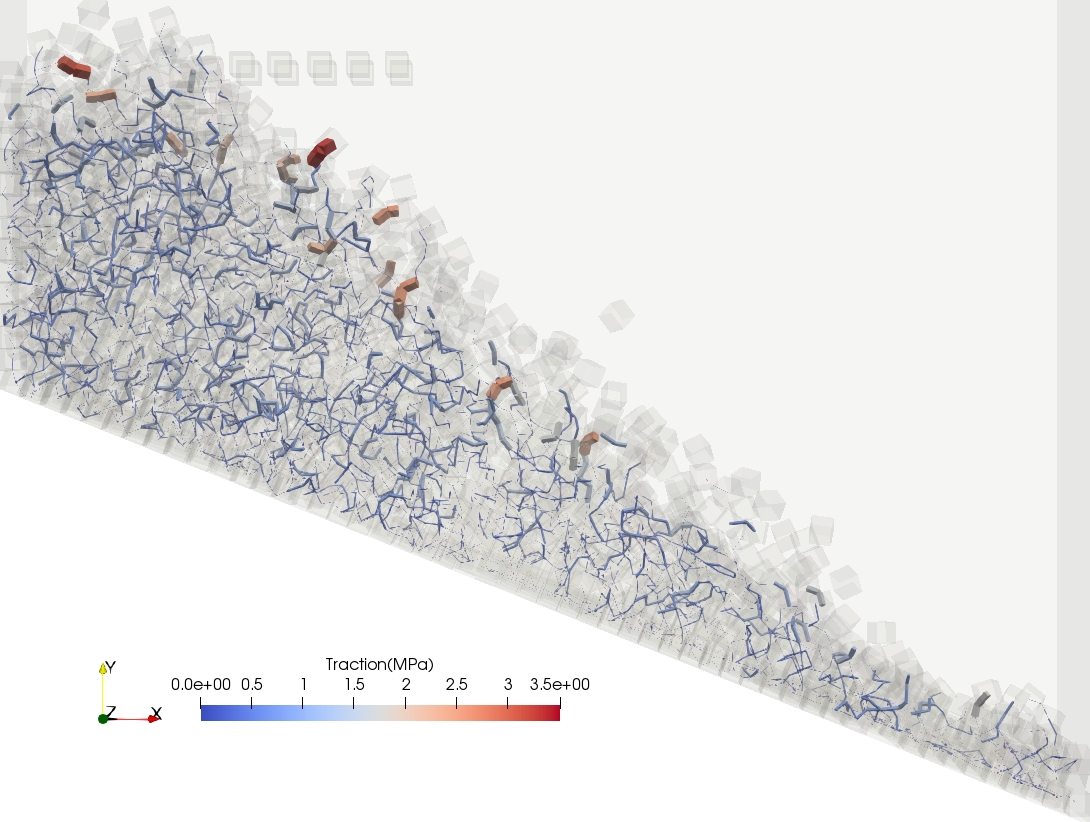}
\par\end{centering}
\begin{centering}
\includegraphics[width=0.33\textwidth]{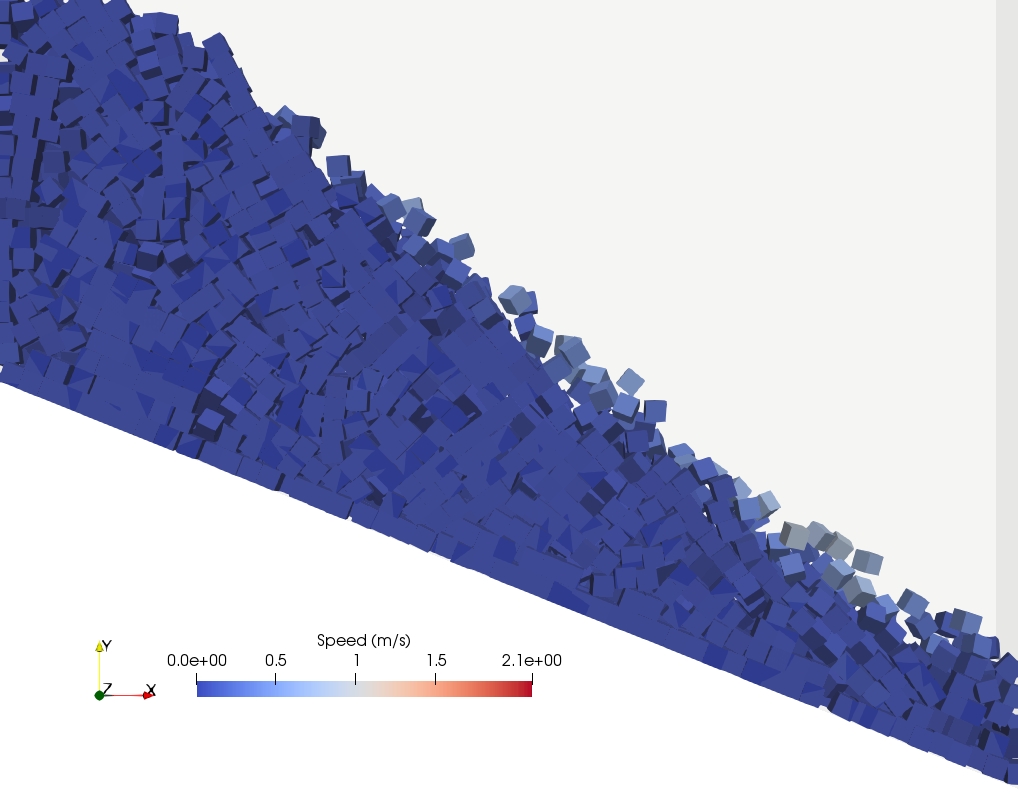}\includegraphics[width=0.33\textwidth]{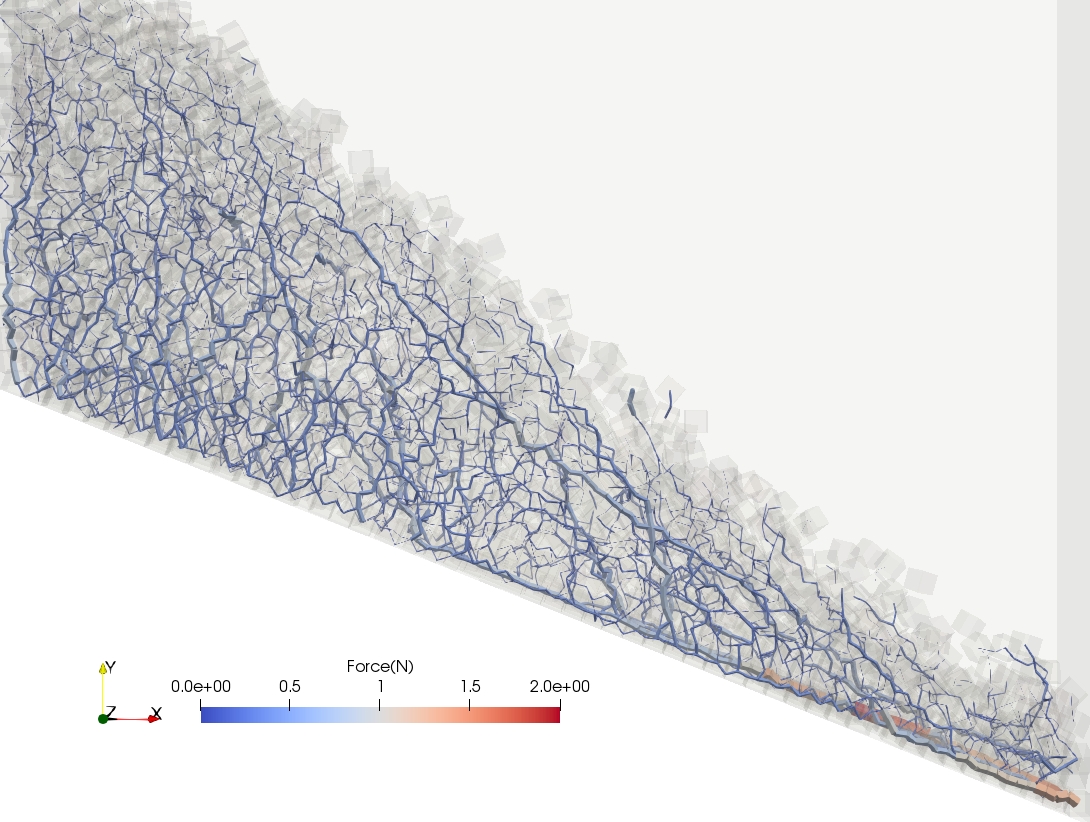}\includegraphics[width=0.33\textwidth]{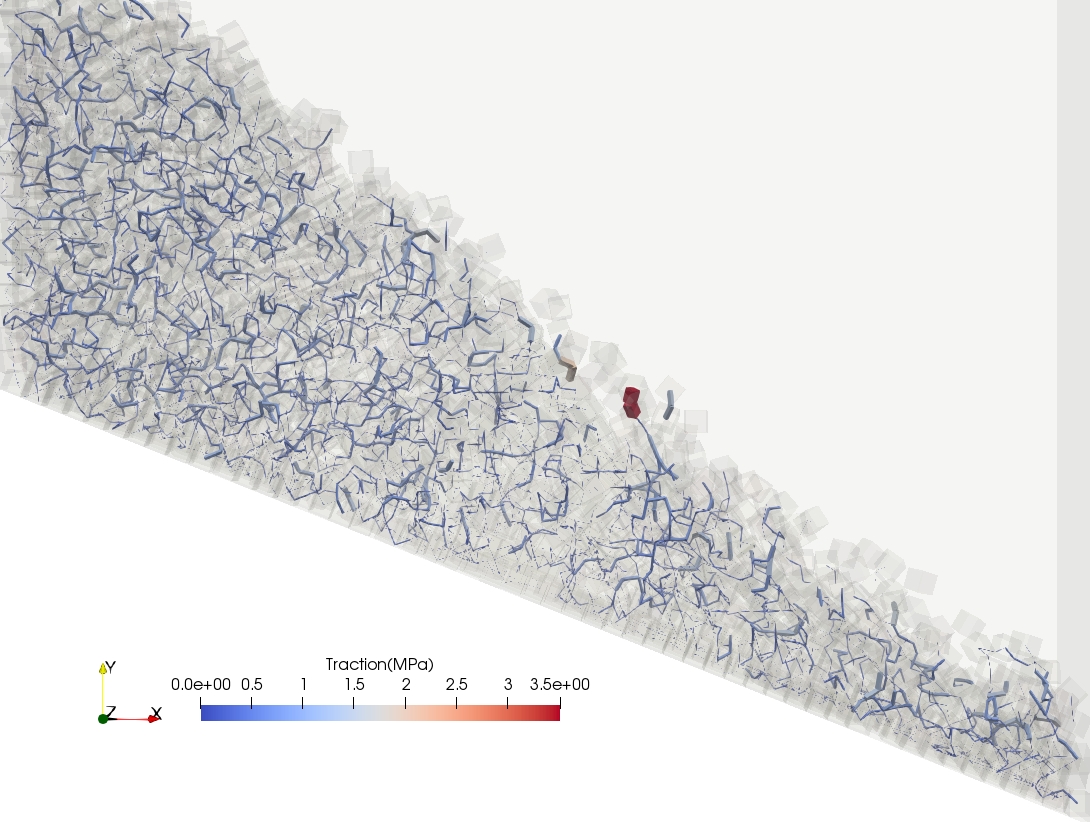}
\par\end{centering}
\begin{centering}
\includegraphics[width=0.33\textwidth]{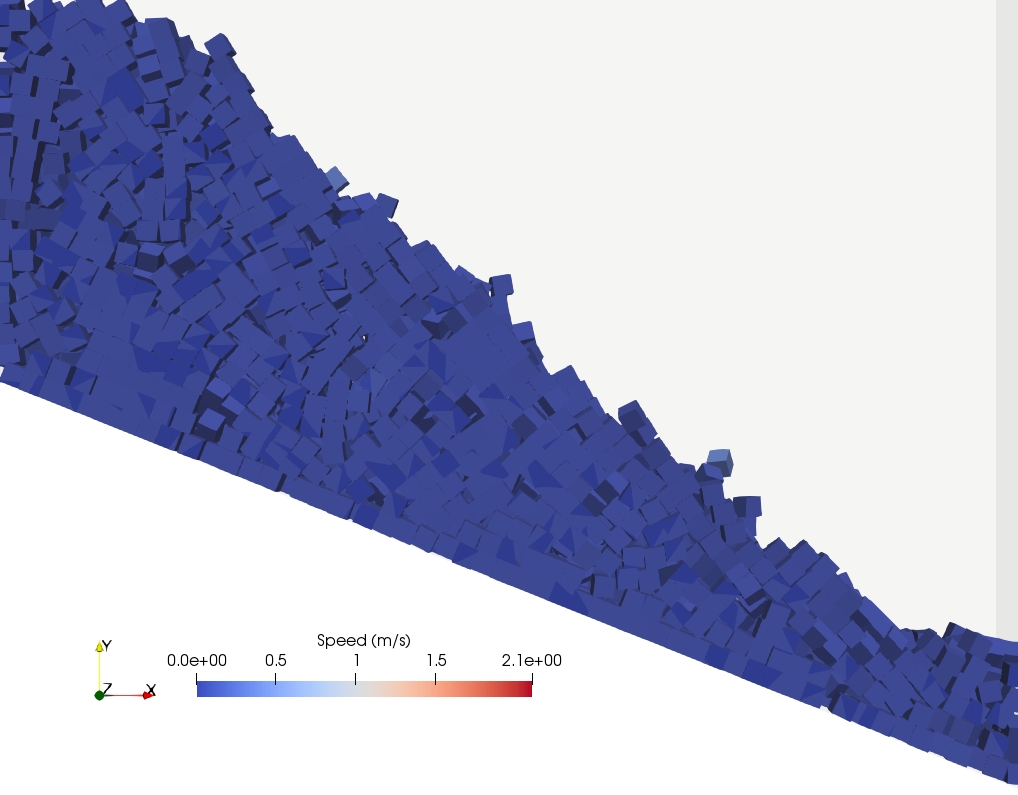}\includegraphics[width=0.33\textwidth]{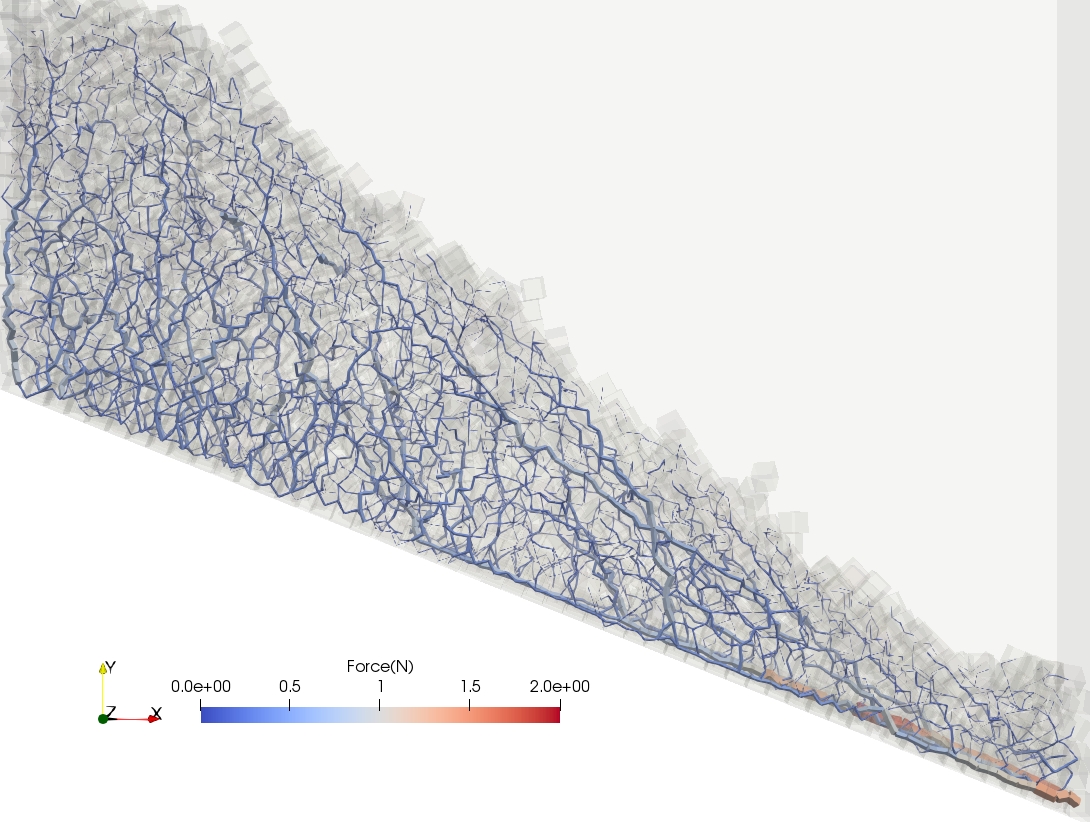}\includegraphics[width=0.33\textwidth]{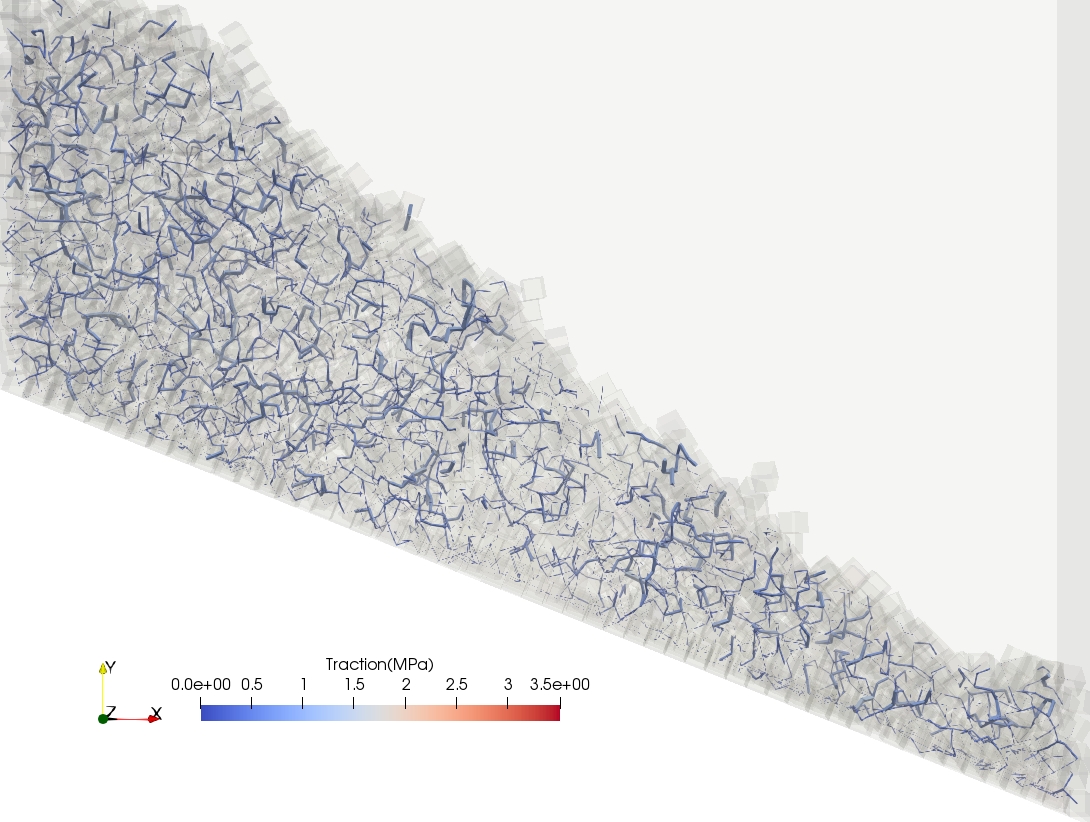}
\par\end{centering}
\caption{Slide box: Speed (left), force chain network (middle) and traction
chain network (right) for cubic particles after 1, 2, 3, 4 and 5 seconds,
with respectively around 680, 1150, 1880, 2000 and 2000 particles,
from top to bottom.\label{fig:tiltboxtime}}
\end{figure}

Next we consider 2000 cubic particles weighing 0.0196 N each, i,.e.
a side length 10mm loaded into the slide box depicted in Figure \ref{fig:tiltbox-setup}.
The simulation after 1s, 2s, 3s, 4s, and 5s are shown in Figure \ref{fig:tiltboxtime}.

The particle-wall static and kinetic coefficients of friction are
0.1, the coefficient of restitution is 0.1. The particle-wall and
particle-particle normal and tangential stiffnesses are 5000 $N.m^{-3}$
and 2500 $N.m^{-3}$, respectively. The particle-wall and particle-particle
rolling damping is 0.1. The particle-particle kinetic coefficients
of friction are 0.4, the coefficient of restitution is 0.4. The simulation
is conducted with a modified inertia tensor.

The largest forces in the force chain network are spatially distinct
from the largest tractions in the traction chain network. Larger forces
tend to build up in the bottom right corner of the slide box. However,
reorientation of the cubic particles allows for face-face contact
between particles, which reduces mean inter-particle tractions. Larger
tractions tend to be isolated towards the top surface due to dynamic
particle-particle interactions over edges. This demonstrates the complementary
nature of the information contained in the two networks. For non-spherical
particle systems, the force chain network only gives partial insight
into inter-particle contact. Thus, the traction chain network complements
this to give additional insight into the inter-particle contact revealing
contact with high mean inter-particle tractions (mean inter-particle
force intensity or mean inter-particle pressure). Similarly, the traction
chain network in isolation only gives partial insight. This highlights
the importance of considering both force chain and traction chain
networks when interpreting non-spherical granular systems.

\section{Conclusion\label{sec:Conclusion}}

Force chain networks have been a cornerstone of quantitative granular
material analysis, offering significant insight into inter-particle
interactions through the network topology and force network evolution.
In addition, the correlation between penetration distance, elastic
contact force magnitude and contact area for conservative spherical
particle systems render force chain networks sufficient and information
complete.

Recent advances in shape accurate discrete element modeling demand
a revisit of force chain networks. In this study, I first demonstrated
that the elastic force magnitude and contact area for non-conservative
spherical particle systems and non-spherical particle systems can
be uncorrelated. The implication of this is that the force chain network
is distinct from the traction chain network, with each contributing
insight into the state of inter-particle contact. 

The traction chain network quantifies the network topology of the
tractions that exist between interacting particles, particularly when
non-spherical and polyhedral shaped particles are considered. Thus,
traction chain networks complement force chain networks with additional
insight into the mean tractions or mean force intensity (or contact
pressures) between interacting particles.

As the additional insights gained from traction chain networks mature,
it will complement contact and particle degradation research for non-spherical
particle systems.

\section*{Acknowledgments\label{sec:Acknowledgments}}

Nvidia for sponsoring GPU resources to conduct this research is acknowledged.

\newpage{}

\bibliographystyle{elsarticle-num}
\bibliography{bibtex_databasem}

\end{document}